\documentclass[fleqn,usenatbib]{mnras}

\usepackage{newtxtext,newtxmath}

\usepackage[T1]{fontenc}
\usepackage{ae,aecompl}
\usepackage[toc,page]{appendix}

\usepackage{graphicx}	
\usepackage{amsmath}	
\usepackage{amssymb}	

\title[The number of globular clusters around DF44]{The number of globular clusters around the iconic UDG DF44 is as expected for dwarf galaxies}

\author[Teymoor Saifollahi et al.]{Teymoor Saifollahi$^{1}$\thanks{E-mail: teymur.saif@gmail.com},  
Ignacio Trujillo$^{2,3}$, 
Michael A. Beasley$^{2,3}$, 
Reynier F. Peletier$^{1}$,
\newauthor
Johan H. Knapen$^{2,3}$
\\
$^{1}$Kapteyn Astronomical Institute, University of Groningen, the Netherlands\\
$^{2}$Instituto de Astrof\'isica de Canarias, V\'ia L\'actea S/N, E-38205 La Laguna, Spain\\
$^{3}$Departamento de Astrof\'isica, Universidad de La Laguna, E-38206 La Laguna, Spain\\
}
\date{Accepted XXX. Received YYY; in original form ZZZ}
\pubyear{2019}
\begin{document}
\label{firstpage}
\pagerange{\pageref{firstpage}--\pageref{lastpage}}
\maketitle
\begin{abstract}
There is a growing consensus that the vast majority of ultra-diffuse galaxies (UDGs) are dwarf galaxies. However, there remain a few UDGs that seem to be special in terms of their globular cluster (GC) systems. In particular, according to some authors, certain UDGs exhibit large GC populations when compared to expectations from their stellar (or total) mass. Among these special UDGs, DF44 in the Coma cluster is one of the better-known examples. DF44 has been claimed to have a relatively high number of GCs,  $N_{GC}=74^{+18}_{-18}$, for a stellar mass of only $3\times 10^8$ $M_{ \odot }$ which would indicate a much larger dark halo mass than dwarfs of similar stellar mass. In this paper we revisit this number and, contrary to previous results, find $N_{GC}=21^{+7}_{-9}$ assuming that the distribution of the GCs follows the same geometry as the galaxy. If we assume that the GCs around DF44 are distributed in a (projected) circularly symmetric way and, if we use a less strict criterion for the selection of the GCs, we find $N_{GC}=18^{+23}_{-12}$. Making use of the $M_{\rm GC} - M_{\rm halo}$ relation, this number of GCs suggests a dark matter halo mass of $M_{halo}=1.1^{+0.4}_{-0.5} \times 10^{11} M_{\odot}$, a value which is consistent with the expected total mass for DF44 based on its velocity dispersion,  $\sigma=33^{+3}_{-3}$ km s$^{-1}$. We conclude that the number of GCs around DF44 is as expected for regular dwarf galaxies of similar stellar mass and DF44 is not extraordinary in this respect.
\end{abstract}




\section{Introduction}

The nature of very low surface brightness galaxies with large effective radii has been intensely debated in past years. Such objects were discovered in the 80s \citep{Binggeli,impey,Dalcanton}, but with new detailed observations have recently been dubbed "Ultra Diffuse Galaxies" \citep[UDGs;][]{vd15}. The debate has focussed on the possible differences between UDGs and the general galaxy population with the same luminosity (i.e. dwarf galaxies\footnote{Historically, the term dwarf galaxy has referred to those galaxies that have a low total luminosity and a low central surface brightness in the $\mu_0$-magnitude plane \citep[see an extended discussion on][]{Binggeli1994}. This terminology has been independent of a galaxy's extension and does not include the compact dwarf ellipticals (like M32). In this sense, an alternative name to UDGs would be “large dwarfs”, as already suggested in the 1980s \citep[][]{Sandage1984}.}) and, in particular, on the amount of dark matter these galaxies may possess.
Are UDGs "normal" dwarf galaxies with relative little star formation activity in their central regions? Or, are UDGs a new type of galaxy with either surprisingly large amounts of dark matter for their stellar mass \citep[see e.g.][]{vd16} or, on the contrary, very little dark matter \citep[see e.g.][]{2018Natur.555..629V,2019MNRAS.486.1192T,Emsellem2019,oliver2020}? 

The vast majority of works point to UDGs having the properties of dwarf galaxies \citep[see e.g.][]{beasley2016,javier,venhola2017,ruizlara2018,pavel2019,fensch2019} but with  flatter light distributions \citep{chamba,nacho2020}. This flatter light distribution could be caused either by tidal interactions \citep{collins2013,Rong2019}, higher internal angular momentum \citep{amorisco2016,pavel2020} or outflows \citep{cintio2017}. However, some UDGs do not fit nicely into the categories of "normal" dwarf galaxies, as their dark matter content has been suggested to be very high. In particular, one of the better-known examples of such an extreme UDG is Dragonfly 44 \citep[DF44; ][]{vd16}. This iconic galaxy, associated with the Coma galaxy cluster, has been claimed to have a dark matter halo comparable with that measured for the Milky Way \citep{vd16}. 

The first study of DF44 \citep[][henceforth vD16]{vd16} measured a high stellar velocity dispersion $\sigma$=47$^{+8}_{-6}$ km s$^{-1}$ within the effective radius of the galaxy. By comparing to theoretical NFW profiles, vD16  suggested that DF44 harbours a dark matter halo as massive as 10$^{12}$ M$_{ \odot }$. Considering the DF44 luminosity M$_V$=-16.2 mag and a stellar mass $M_{*}=3 \times 10^8 M_{ \odot }$, the authors of this paper claimed that DF44 resembles a 'failed' Milky Way. 

This conclusion was further supported by the assertion of an extensive number of globular clusters (GCs) in the vicinity of the galaxy \citep[][henceforth vD17]{vd17}. In the absence of a dynamical estimation of the dark matter halo mass based on the kinematics of the GCs around the galaxy, the GC number count (N$_{GC}$) can be used as a good proxy  for regular galaxies \citep{Spitler2009,harris2013,Hudson2014} and UDGs \citep{beasley2016,beasley2016b,peng2016,harris2017,lim2018,amorisco2018,udgngc3,udgngc4}. 

Using HST data, vD17 measured N$_{GC}$=74$^{+18}_{-18}$ GCs around DF44 which implies a mass for the dark matter halo of M$_{halo}$=5$\times$10$^{11}$ M$_{\odot}$ using the GC system mass -- halo mass relations. This number decreases the original claim by the same group of N$_{GC}$=94$^{+25}_{-20}$ (vD16) based on ground-based data.  Later, \citet[][henceforth vD19]{vd19}, using spatially resolved stellar kinematics, reported a smaller velocity dispersion for DF44 ($\sigma$=33$^{+3}_{-3}$ km s$^{-1}$, vD19) compared to the previous estimation ($\sigma$=47$^{+8}_{-6}$ km s$^{-1}$, vD16). Using the new velocity dispersion, vD19 decreased by an almost an order of magnitude the amount of dark matter in the halo of DF44, from M$_{halo}$=10$^{12}$ M$_{\odot}$ and M/L(<r$_{1/2}$)=48$^{+21}_{-14}$ to  M$_{halo}$=1.6$\times$10$^{11}$ M$_{\odot}$ and M/L(<r$_{1/2}$)=26$^{+7}_{-6}$. However, the large number of GCs ($\sim$75) remains in strong tension with the expected number according to the stellar mass of the object (we would expect $\sim$20 GCs for the stellar mass of DF44, if this galaxy follows the stellar mass -- halo mass relation). 

In this paper, we revisit the GC population of DF44 by exploring their spatial distribution, luminosity function, number count and average colour and we find a significantly lower number of GCs, making this galaxy appear more consistent with the general dwarf galaxy population. In this work, the distance to DF44 (100 Mpc), its  absolute magnitude (M$_V$=-16.2 mag) and velocity dispersion ($\sigma=33^{+3}_{-3}$ km s$^{-1}$) are from vD17 and vD19. All the magnitudes and colours are expressed in the AB magnitude system (unless explicitly stated otherwise).  Throughout this paper, we assumed  the cosmological model with $\Omega_{M}=0.3$, $\Omega_{\Lambda}=0.7$ and H$_{0}$=70 km s$^{-1}$ Mpc$^{-1}$.

\section{Data}

DF44 imaging data was retrieved from the Hubble Space Telescope archive (HST Proposal 14643, PI: van Dokkum). This is the same dataset used by vD17. The data comprise three orbits in \textsl{V$_{606}$} with exposures ranging from 2200-2400s and one orbit in \textsl{I$_{814}$} with 2200s exposure time. \textsl{V$_{606}$} images were median combined using \textit{SWarp} \citep{swarp} and the final amount of time in this band is 7280s. The depth of the images for point sources (5$\sigma$) are 
 \textsl{V$_{606}$}=28.4 mag and \textsl{I$_{814}$}=26.8 mag. 
 
\begin{figure*}
\centering
\includegraphics[trim=90 70 70 70, clip,width=0.9\linewidth]{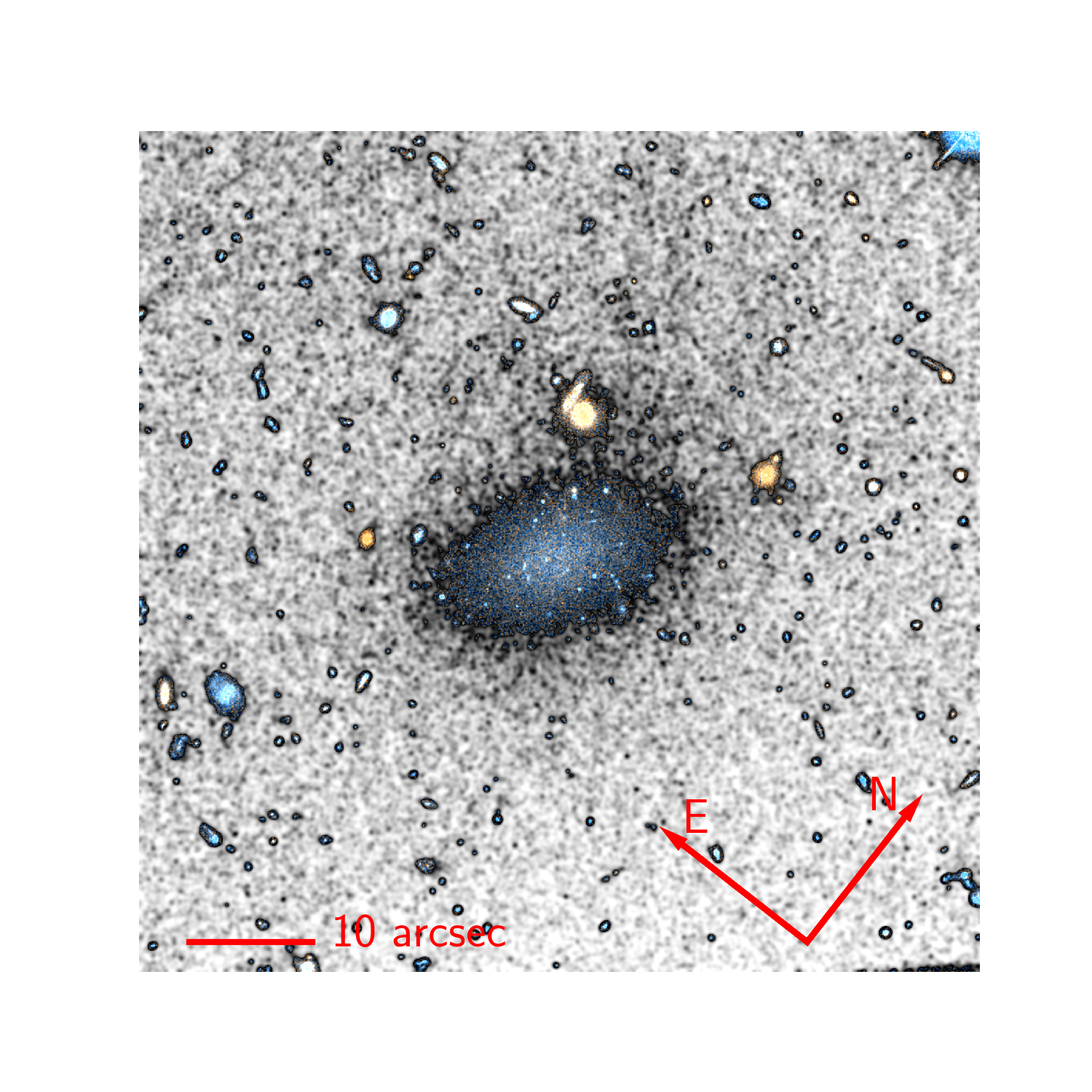}
\caption{Colour composite image of DF44 combining \textsl{V$_{606}$} and \textsl{I$_{814}$} filters. The black and white background corresponds to the \textsl{V$_{606}$} filter. The surface brightness limit of the image is $\sim$28.5 mag/arcsec$^2$ (3$\sigma$; 3\arcsec$\times$3\arcsec).}
\label{figure-imagedf44}
\end{figure*}

 The field of view of the WFC3 is 162\arcsec$\times$162\arcsec and pixel size is 0.0396\arcsec. Instrumental zero-points (AB/mag) were calculated using the \textsl{PHOTFLAM} and \textsl{PHOTPLAM} values as given in the data analysis handbook of the instrument \citep{wfc3}. The calculated zeropoint values for \textsl{V$_{606}$} and \textsl{I$_{814}$} are 26.10 mag and 25.14 mag respectively. Fig. \ref{figure-imagedf44} shows a colour composite image of DF44 combining \textsl{V$_{606}$} and \textsl{I$_{814}$} filters.

\section{Analysis}

\subsection{Structural parameters of the DF44 galaxy}

\begin{figure} 
\centering
\includegraphics[width=\linewidth]{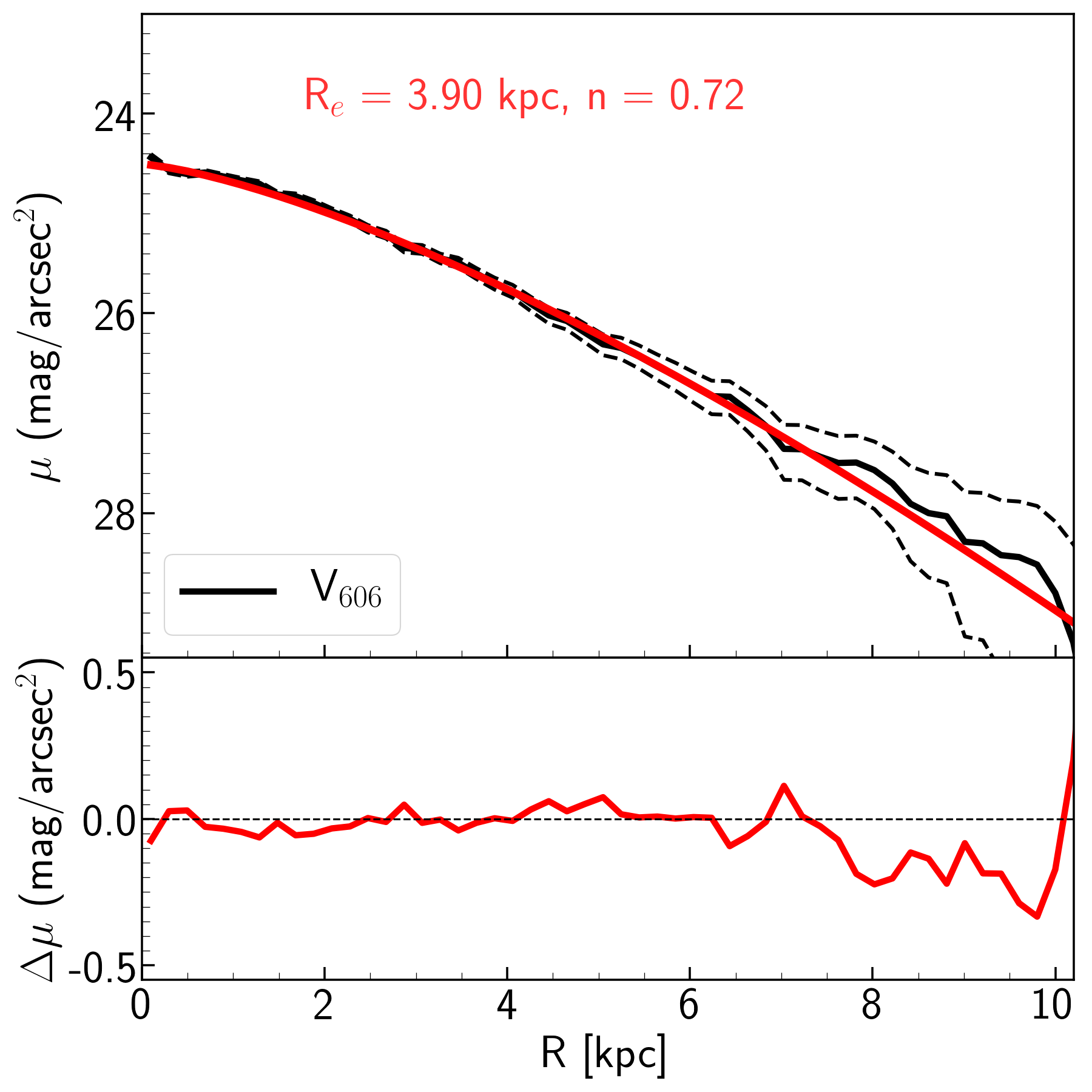}
\caption{Surface brightness profile of DF44 in the \textsl{V$_{606}$} band. Together with the observed profile (black solid line), we also include the best fitting (red curve) using a S\'ersic model (R$_e$=3.9$\pm$0.7 kpc and n=0.72$\pm$0.14). The dashed lines indicate the $\pm$1$\sigma$ uncertainty on  the surface brightness profile. The lower panel shows the residuals from the fits.}
\label{df44lightprofile}
\end{figure}

In order to extract the structural properties of DF44, we have used the code IMFIT \citep{2015ApJ...799..226E}. We assumed, as vD17 did, that the galaxy is well described by a S\'ersic model. The model was convolved with the point spread function (PSF) of the HST image. To conduct a proper fit, we have masked the background galaxies and the foreground stars in the images. Fig. \ref{df44lightprofile} shows the observed light profile of the galaxy in the \textsl{V$_{606}$} band and the corresponding S\'ersic fit. The structural properties of the galaxy are R$_e$=3.9$\pm$0.7 kpc and S\'ersic index n=0.72$\pm$0.14. The axis ratio is q=0.66$\pm$0.01 and the position angle PA=-26.4$\pm$0.7 (measured from North to East, counterclockwise). 

The various published values of effective radius, R$_e$, of DF44 have been estimated using different data and assumptions about the shape of its surface brightness profile. For instance, \citet{vd15} used CFHT imaging and assumed an exponential surface brightness profile for DF44 to get R$_e$=4.6$^{+1.5}_{-0.8}$ kpc. Using  Keck deep imaging, \citet{vd15b} found R$_e$=4.3$\pm$0.3 kpc using a single S\'ersic component with n=0.89$\pm$0.06 and R$_e$=4.1 kpc when two components are used for fitting DF44. Later, using Gemini deep data, \citet{vd16} measured R$_e$=4.3$\pm$0.2 kpc fitting a S\'ersic model with n=0.85. Using the same data but this time calculating R$_e$ from the growth curve, \citet{chamba} got R$_e$=3.3$\pm$0.3 kpc. Finally, vD17 using the same HST dataset than the one we have used here found R$_{e,vD17}$=4.7 kpc and n$_{vD17}$=0.94\footnote{Uncertainties of R$_{e}$ and n are not provided in vD17.}. A possible explanation for the different values of R$_e$ measured here and in vD17 involves the estimation of the local background around the galaxy, a slight change in which affects the determination of n, and ultimately, the value of R$_e$.

\subsection{Detection of globular cluster candidates}

\begin{figure}
\centering
\includegraphics[trim=70 60 70 60, clip, width=\linewidth]{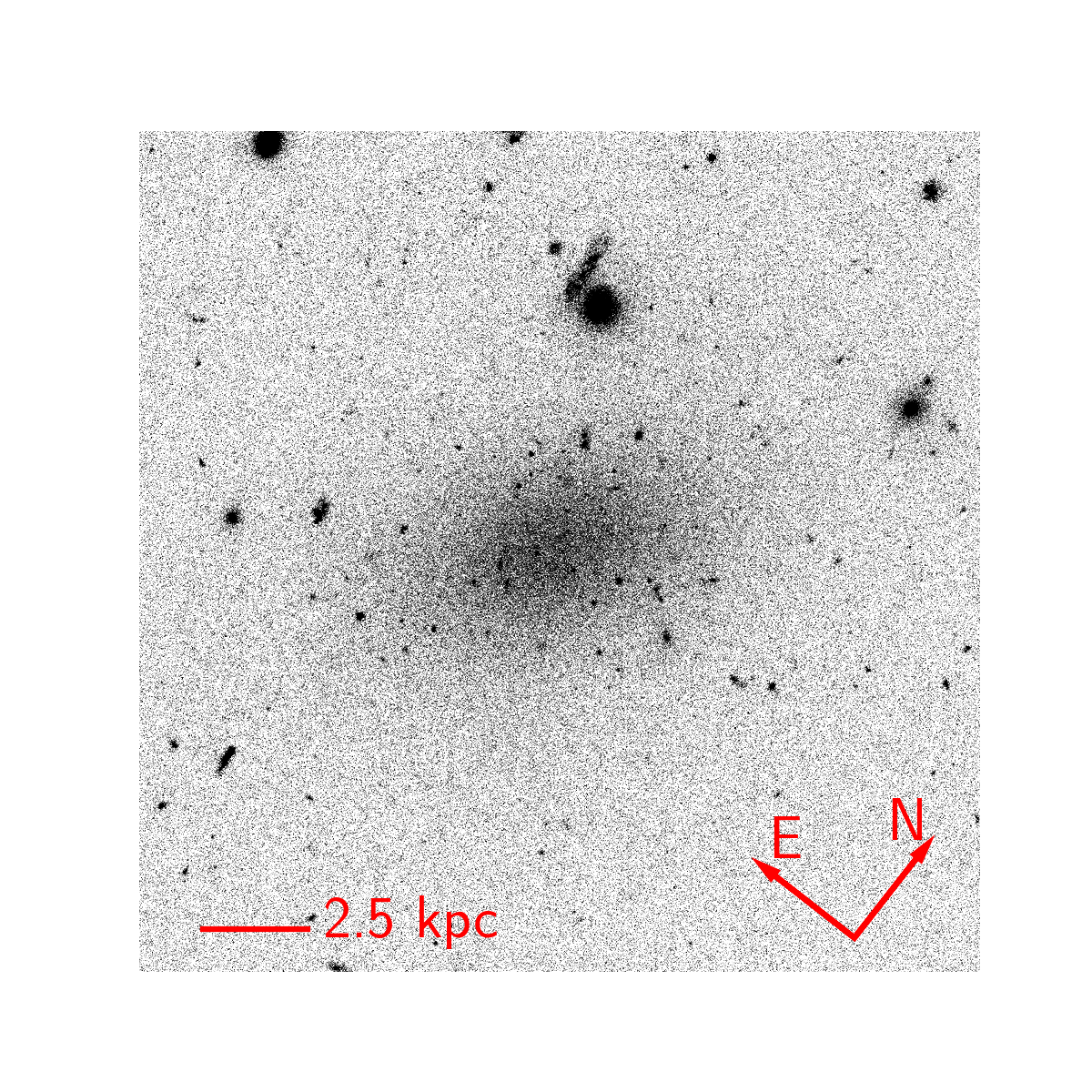}
\includegraphics[trim=70 60 70 60, clip, width=\linewidth]{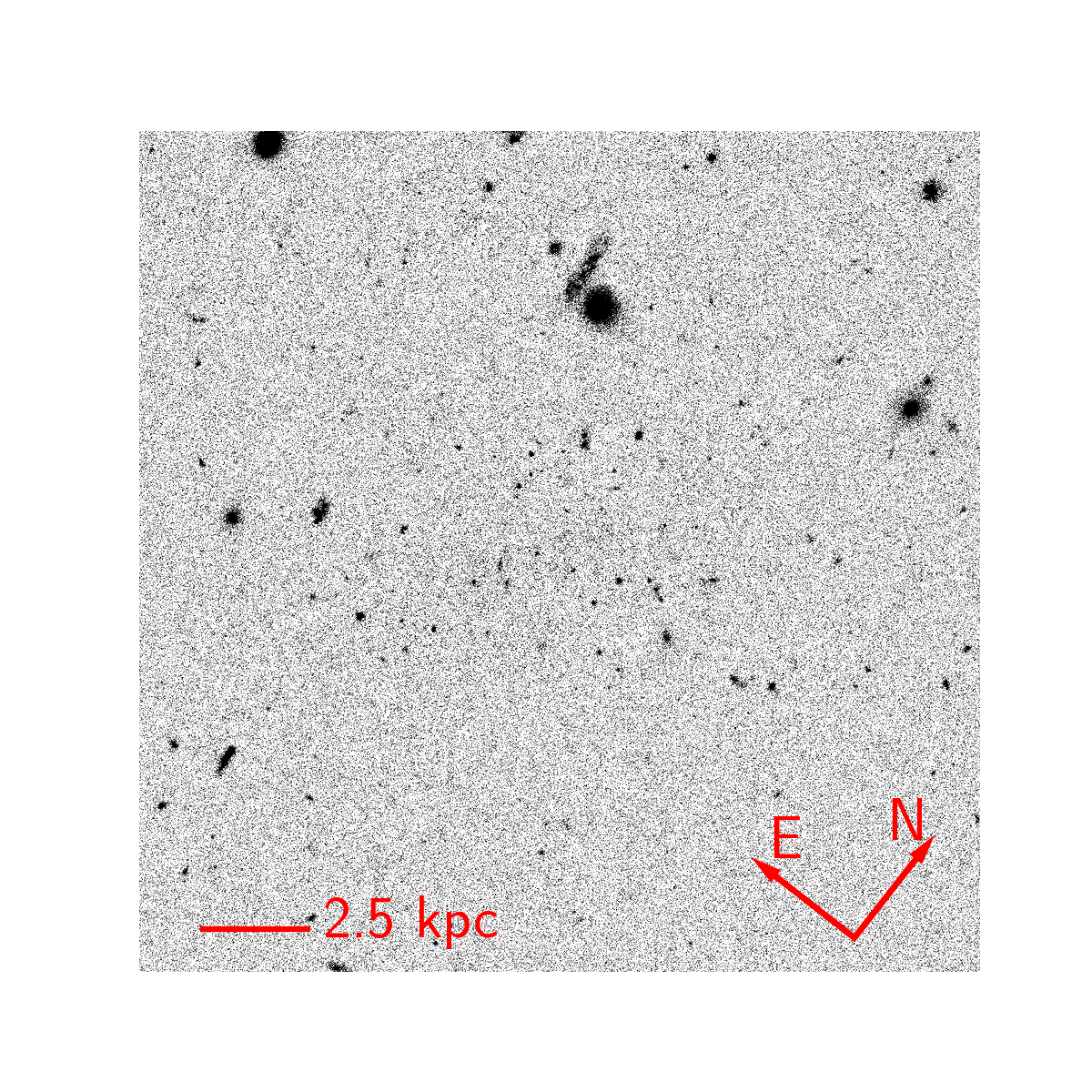}
\caption{\textit{Top}: DF44 as seen by HST using the \textsl{V$_{606}$} filter. \textit{Bottom}: Median filtered (32$\times$32 pixels) removal of the extended light distribution of the galaxy to highlight the presence of point-like sources on the image.}
\label{df44subtracted}
\end{figure}

The detection of the GC candidates around DF44 was performed in the deepest image available: \textsl{V$_{606}$}. \textit{SExtractor} \citep{sex} was used to extract all the sources from the \textsl{V$_{606}$} image to perform aperture photometry. First, a background model was made using a 32$\times$32 pixels median filter and subtracted from the final frame. This background subtraction removes the diffuse light of DF44 and improves the detection efficiency of compact objects in the vicinity of the galaxy (bottom panel in Fig. \ref{df44subtracted}).

Three different techniques were used to explore the effect of the diffuse light subtraction on the catalogues of point-like sources we retrieved from the images, namely :  i. unsharp masking, ii. median filtering and subtracting and iii. S\'ersic fitting and subtracting the galaxy model. These three different methods produce the same catalogues of point-like sources down to \textsl{V$_{606}$}=28.5 mag. We chose the median-filtered approach to assure that the removal of the diffuse light was done homogeneously throughout the entire image.

To characterize the objects in the image, we used  \textit{SExtractor}. Most of the \textit{SExtractor} parameters were left to their default settings except for BACK\_SIZE, DEBLEND\_NTHRESH and DEBLEND\_MINCONT whose values are shown in Table \ref{sextractortable}. To take into account the local variation of the background, including residuals from DF44 diffuse light subtraction, BACK\_SIZE=32 was used (instead of the default value BACK\_SIZE=64). Moreover, to avoid extracting the small features of background galaxies as sources, we adjusted DEBLEND\_NTHRESH=2 and DEBLEND\_MINCONT=0.02 (instead of the default values DEBLEND\_NTHRESH=32 and DEBLEND\_MINCONT=0.005). The same configuration parameters were used to extract sources in the \textsl{I$_{814}$} image.

\begin{table}
\centering
\caption {\textsl{SExtractor} parameters that were applied for source detection and aperture photometry.}
\begin{tabular}{ c c } 
\hline  
Parameter & Value \\
\hline  
DETECT\_MINAREA & 3.0 \\
DETECT\_THRESH & 1.5 \\
ANALYSIS\_THRESH & 1.5 \\
DEBLEND\_NTHRESH & 2 \\
DEBLEND\_MINCONT & 0.02 \\
BACK\_TYPE & GLOBAL \\ 
BACK\_SIZE & 32 \\
BACK\_FILTERSIZE & 3 \\
PHOT\_APERTURE & 4,8,30 \\
\hline
\end{tabular}
\label{sextractortable}
\end{table}

The total magnitudes of the different sources were measured using aperture photometry. We estimated the aperture magnitudes of all the targets using \textit{SExtractor} and PHOT\_APERTURES=4 pixels (diameter). This was done in both bands (\textsl{V$_{606}$} and \textsl{I$_{814}$}) and the magnitudes were aperture-corrected. The aperture correction values are 0.54 and 0.65 mag for \textsl{V$_{606}$} and \textsl{I$_{814}$}, respectively. To estimate such aperture corrections we use a few bright, non-saturated, stars with magnitudes in the \textsl{V$_{606}$} band between 24 and 25.5 mag. For those stars, we calculate two different apertures: PHOT\_APERTURES=4 pixels and 30 pixels. The above aperture corrections mostly corresponds to the amount of light between 4 to 30 pixels. However, beyond 30 pixels there is still some light that is hard to measure using stars with that signal-to-noise. For such reason, we add another 0.07 mag based on a prescription from the WFC3 instrument handbook\footnote{\url{http://documents.stsci.edu/hst/wfc3/documents/handbooks/}}. After this, \textsl{V$_{606}$}-\textsl{I$_{814}$} colours were calculated using the aperture corrected magnitudes. 

At the distance of the Coma cluster (100 Mpc), globular clusters are not resolved \citep{peng2011} and appear as compact as foreground stars. Therefore, to select GCs around DF44, as a first step, we measured the compactness of the detected objects and identified the compact sources. We took the difference between the two aperture magnitudes (of 4 and 8 pixels) to represents the compactness of the objects. In this paper, we denoted the magnitude difference (or compactness parameter) as $\Delta$m$_{4-8}$. Compact objects, compare to more extended objects, display smaller values of $\Delta$m$_{4-8}$. 

To understand how the compactness parameter behaves across the field of view and for different magnitudes, we simulated more than 7500 artificial stars using TinyTim\footnote{\url{http://www.stsci.edu/software/tinytim/}} v7.0. This was done because of the lack of bright stars in the \textsl{V$_{606}$} frame. The artificial stars range in magnitude between 24 and 30. We simulated 128 artificial stars in 60 magnitude bins, randomly distributed on the detector. The TinyTim PSF models take into account many variables such as filter, optical and detector responses with the position on the detector and wavelength, focus, aberrations, geometric distortions and charge diffusion. Moreover, HST long exposures introduce a small displacement (jitter) and change in focus during exposures (berating). For the simulation, we started with jitter=5 mas and applied different values of de-focus between 1 and 10  $\mu$m. The final value of de-focus=6 $\mu$m  was chosen to match the compactness of the artificial stars with the few bright stars observed in the field of view. After producing the artificial stars with TinyTim, Poisson noise was added to each mock star. We finally added them to the main frame (\textsl{V$_{606}$}). 

We explored whether our compactness values can be affected by the effect of the sub-pixel location of the PSFs and by the drizzling algorithm that is used to create the final science images. We found that the subpixel location of real PSFs has a small effect on the compactness parameter ($\sim$0.01 mag). The effect of the drizzling algorithm is the following. When comparing the compactness of real sources in both drizzled and non-drizzled images, we found that the difference in compactness is compatible with zero on average. The scatter between the drizzled and non-drizzled compactness measurements is around 0.1 mag. We conclude that our compactness selection is robust in that sense.

\begin{figure*}
  \includegraphics[width=0.99\linewidth]{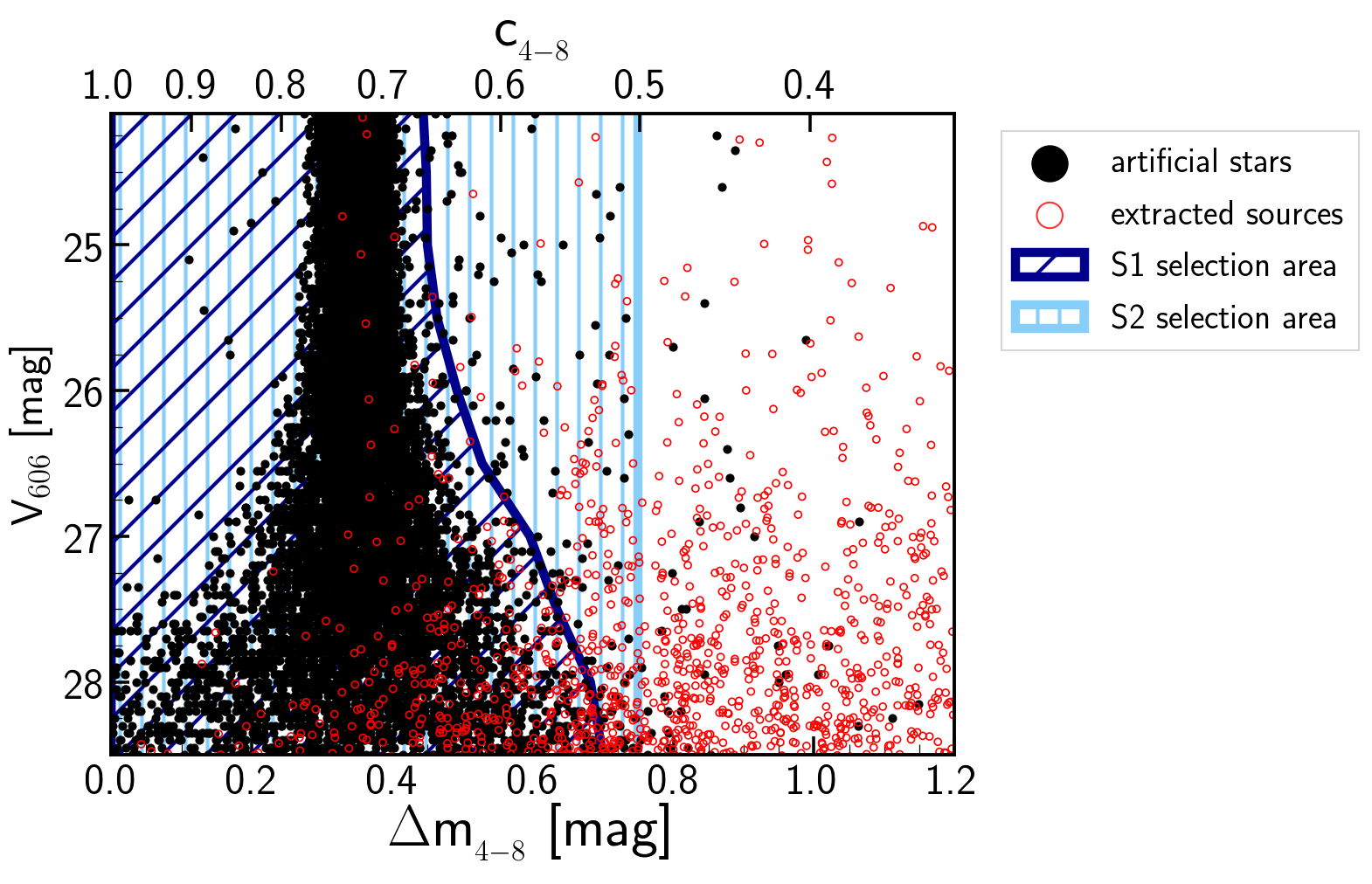}
  \caption{Magnitude (\textsl{V$_{606}$}) versus compactness of the sources $\Delta$m$_{4-8}$ map. This map is used to select the GC candidates in this work. To facilitate for the reader the comparison with the compactness parameter used in vD17, we have included their values in the upper X-axis. Mock point-like sources are shown with black dots, while observed real sources are shown with red dots.
  The compactness parameter $\Delta$m$_{4-8}$ of the brightest artificial stars varies from 0.3 to 0.4, depending on the position on the detector. For the fainter stars, uncertainties in the photometry play an important role and increase the scatter in $\Delta$m$_{4-8}$. Our main sample S1 corresponds to observed objects within the region indicated by the dark blue lines. We have also explored a less restrictive sample of objects S2 (enclosed by the light blue vertical lines, which corresponds to the selection criteria given in vD17).}
  \label{simulation}
\end{figure*}

We found that the compactness parameter $\Delta$m$_{4-8}$  of the artificial stars varies between 0.3 and 0.4 and has an average value of 0.36. This value is consistent with the average compactness of the bright stars in the data. Next, we selected objects with compactness within 3$\sigma$ from the mean values of the artificial stars (for each magnitude bin) as compact sources. Fig. \ref{simulation} shows  $\Delta$m$_{4-8}$ of all the observed sources and the mock stars as a function of their \textsl{V$_{606}$} magnitude. In this work, we have used the region enclosed by the dark blue contour as the selected area for defining our main sample of compact sources (sample S1).

\begin{figure}
\centering
\includegraphics[width=\linewidth]{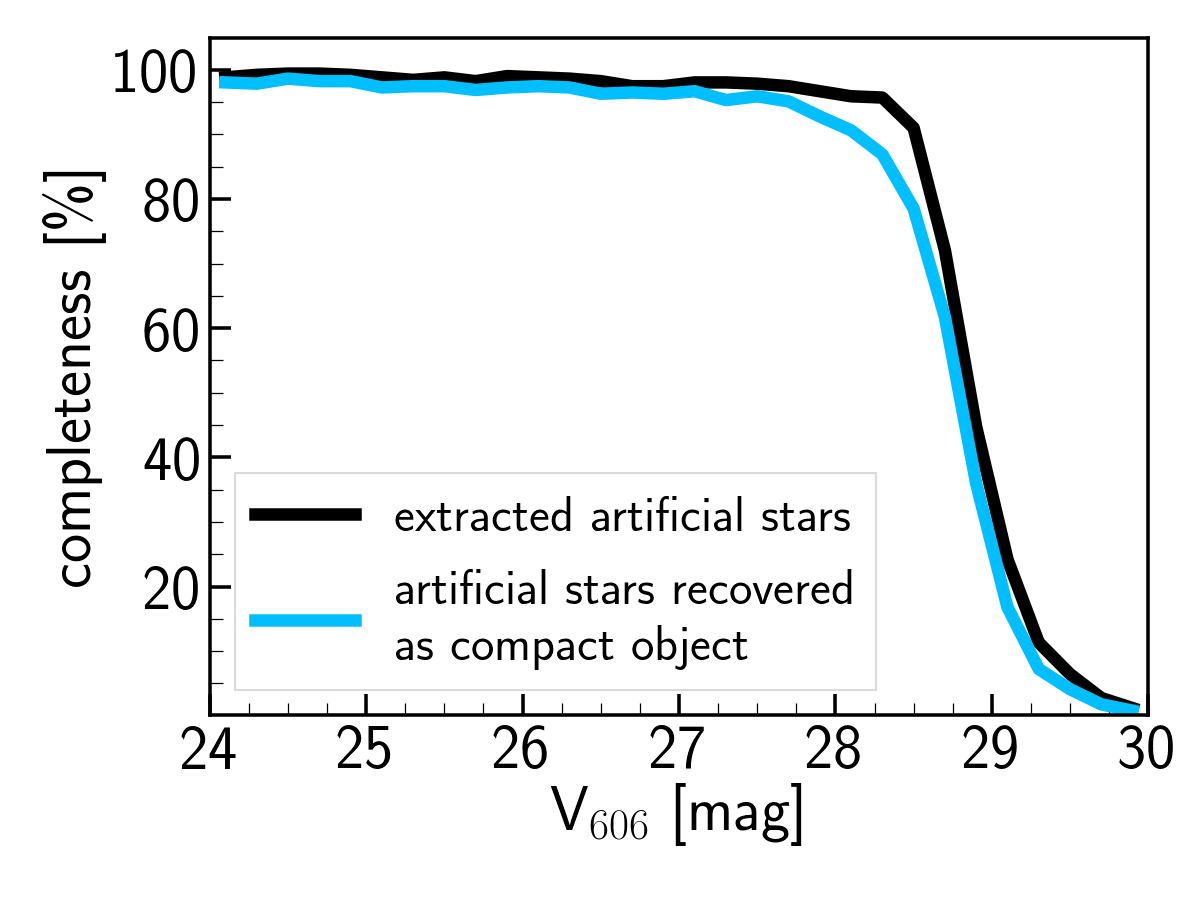}
\caption{The completeness of the selection of the compact sources in S1. This sample is more than 90 percent complete up to magnitude \textsl{V$_{606}$}=28.2 mag. Black line corresponds to the completeness of all the mock objects injected in the image, while the blue line is the fraction that are identified as compact sources with our compactness criteria.}
\label{completeness}
\end{figure}

As is seen in Fig. \ref{completeness},  sample S1 is more than 90\% complete up to magnitude \textsl{V$_{606}$}=28.2 mag and more than 80\% complete to magnitude \textsl{V$_{606}$}=28.5 mag. As a further test to test our selection criteria, we have measured the ellipticities of the objects characterised as compact and extended. We find  the average ellipticity of the brightest compact and extended sources (brighter than $V_{606}$$<$26.0 mag) is 0.06 ($\sigma$=0.04) and 0.33 ($\sigma$=0.2) respectively.

To be able to compare our sample with that of  vD17, we select another sample S2 following their selection criteria 0.5$<$c$_{4-8}$$<$1.0. The compactness parameter c$_{4-8}$ is the flux ratio between two different apertures, 4 pixels and 8 pixels (according to vD17). The corresponding region in the \textsl{V$_{606}$} vs compactness map is also shown in Fig. \ref{simulation}. As can be seen in the figure, the S1 sample excludes a fraction of sources that are included in S2.

\subsection{The distribution of globular clusters around DF44}

\begin{figure}
\centering
\includegraphics[width=\linewidth]{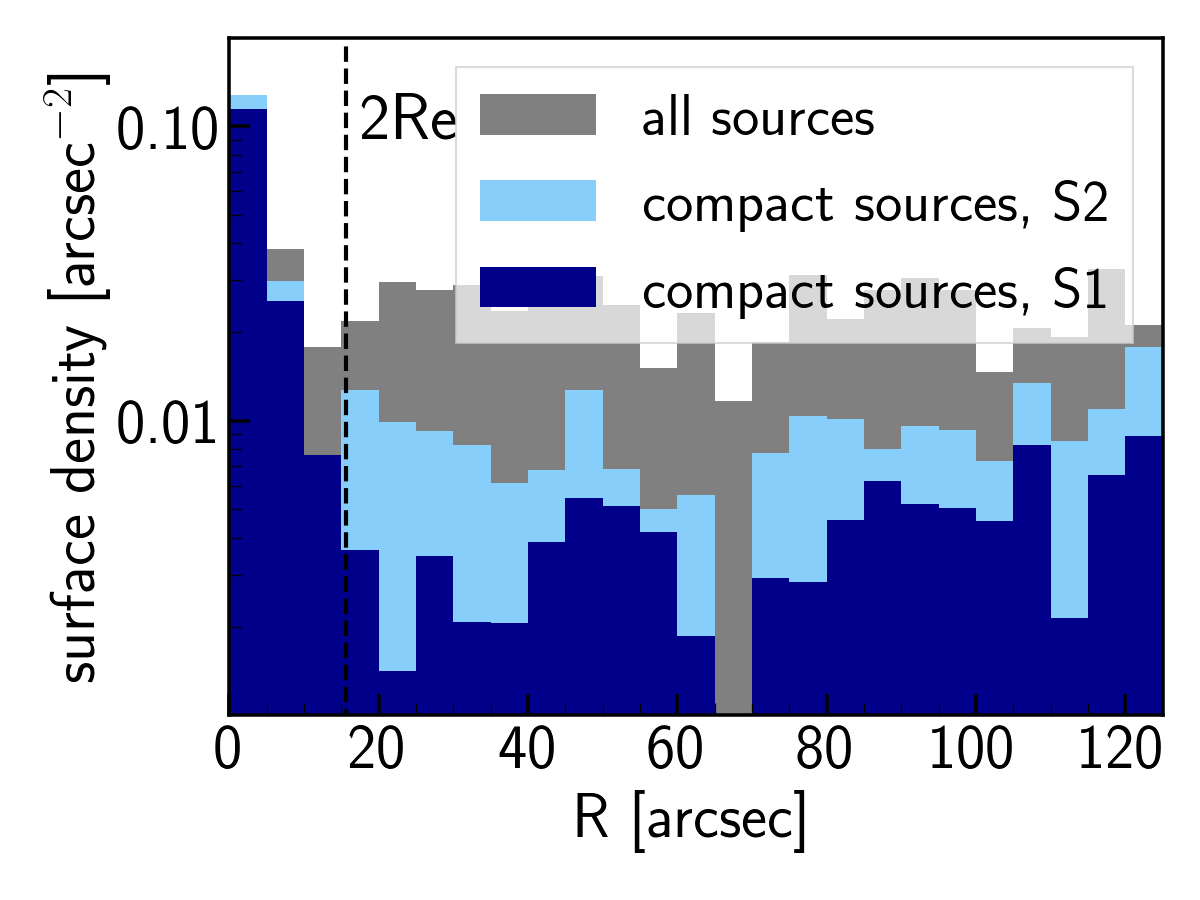}
\caption{Surface density radial distribution of all the extracted sources (grey) around DF44, the compact sources selected in this work S1 (dark blue) and the compact sources selected in vD17 (light blue). We only show sources brighter than \textsl{V$_{606}$}=28.0 mag. The sample selected in this paper S1 is less affected by background contamination than the one following the criteria by vD17.}
\label{radialdistributiongcs}
\end{figure}

Once the selection of the compact sources of the image has been performed, we can explore how they are distributed around DF44. The surface density of the radial distribution of the compact sources is shown in Fig. \ref{radialdistributiongcs}. Here, we only show sources with \textsl{V$_{606}$}$<$28.0 mag. We plotted both the distributions of the compact sources selected in our main sample S1, and the equivalent sample done by vD17, S2. At a radial distance of equivalent to 2R$_e$, both distributions detect a similar number of objects. However, due to the more relaxed selection criteria by vD17, beyond that distance the background density is much more important in S2 than in S1.

\begin{figure*}
    \centering
    \begin{tabular}{l l}
        \textbf{i. S1 sample, GC axis ratio = 1} & \textbf{ii. S1 sample, GC axis ratio = 0.66} \\\\
        \includegraphics[width=0.4\linewidth]{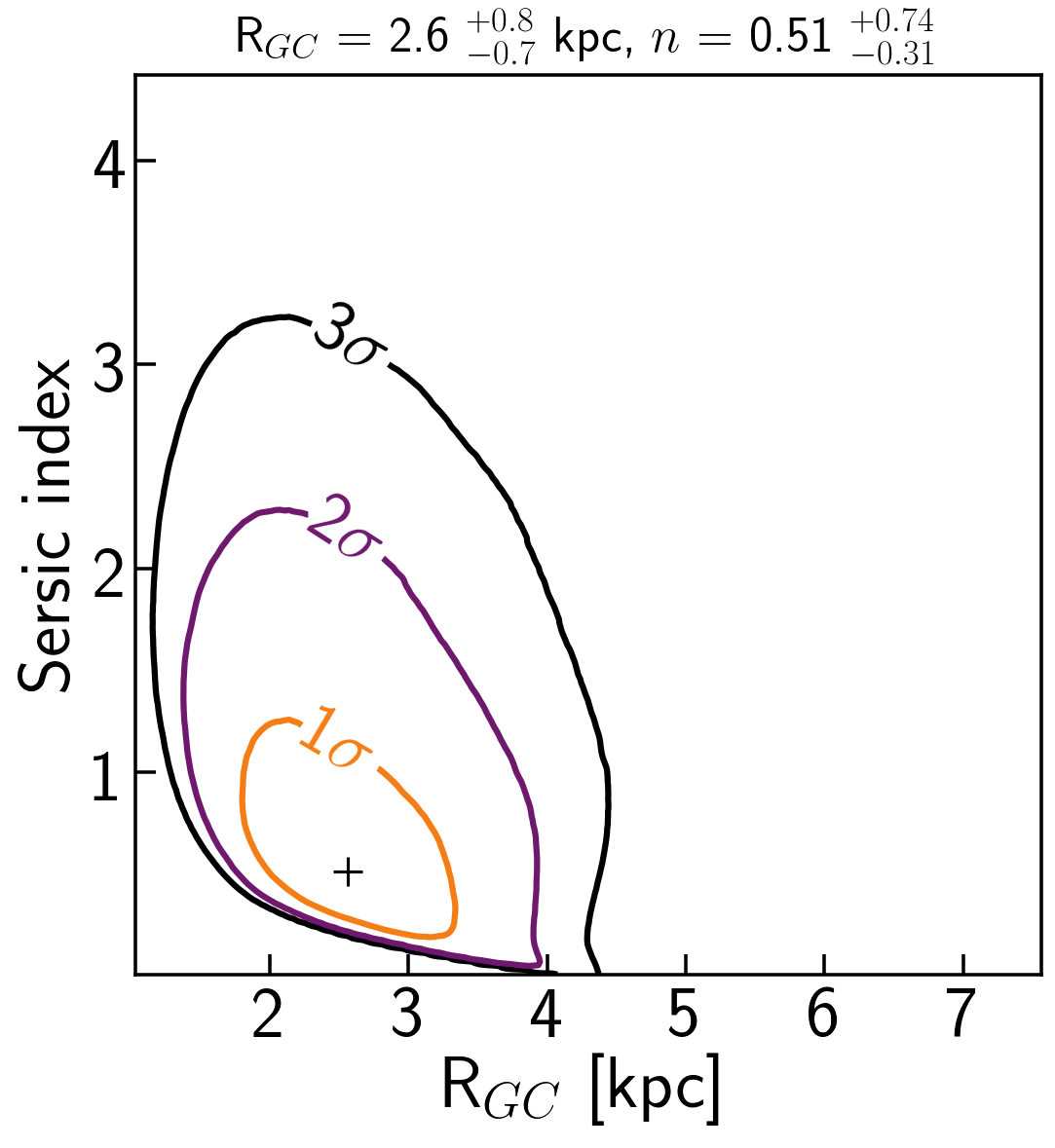} & \includegraphics[width=0.4\linewidth]{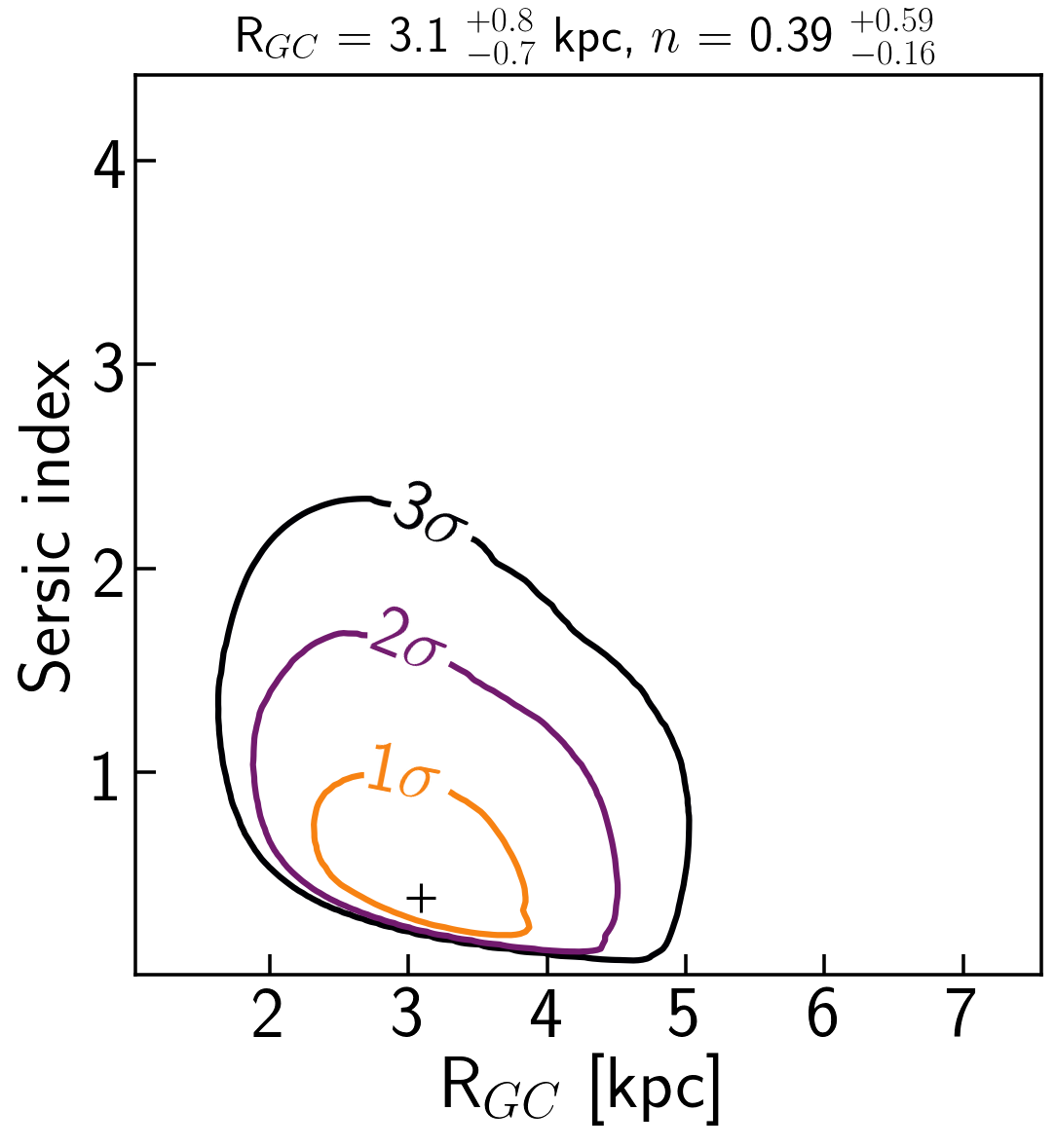} \\\\\\
        \textbf{iii. S2 sample, GC axis ratio = 1} & \textbf{iv. S2 sample, GC axis ratio = 0.66} \\\\
        \includegraphics[width=0.4\linewidth]{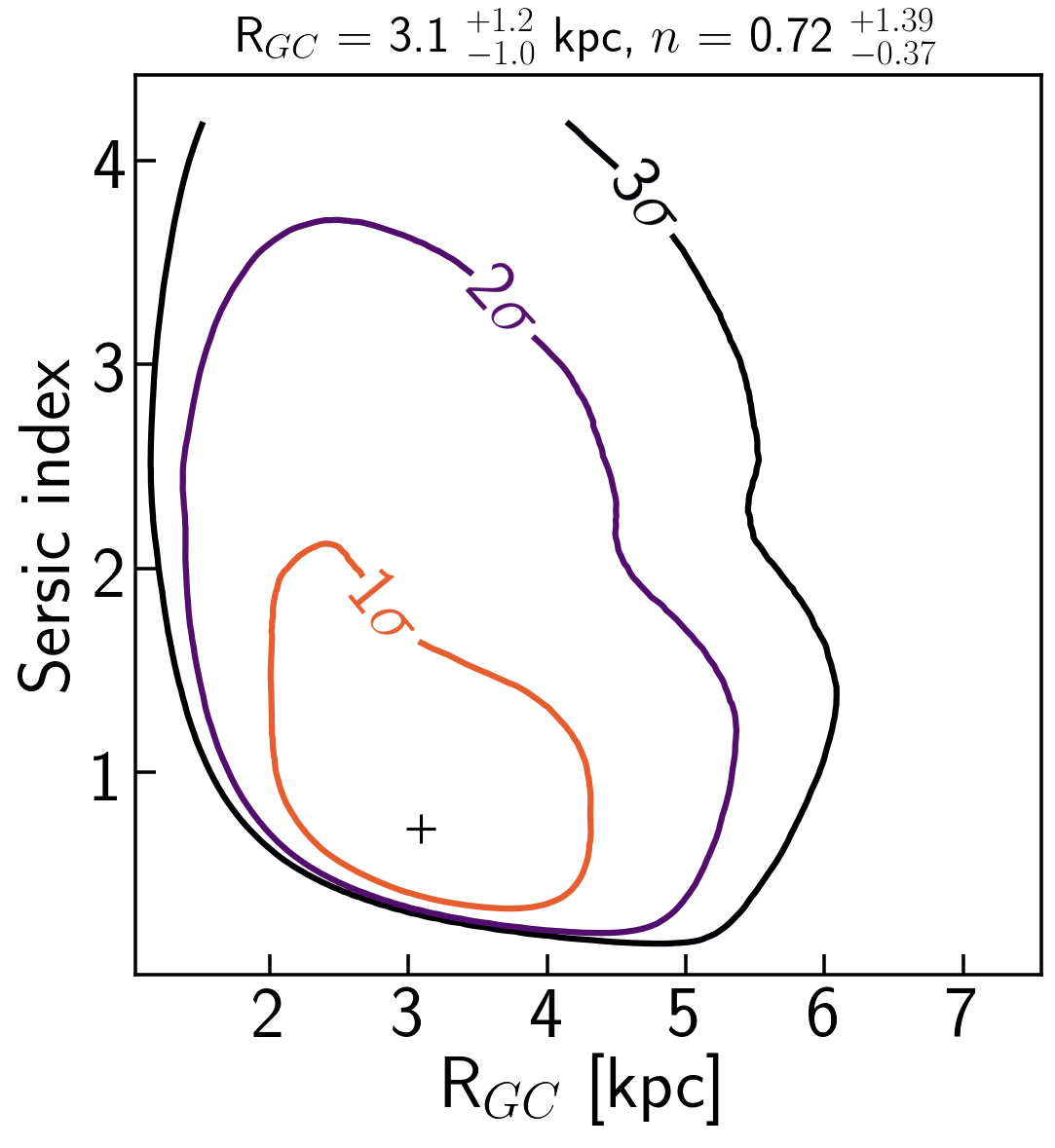} & \includegraphics[width=0.4\linewidth]{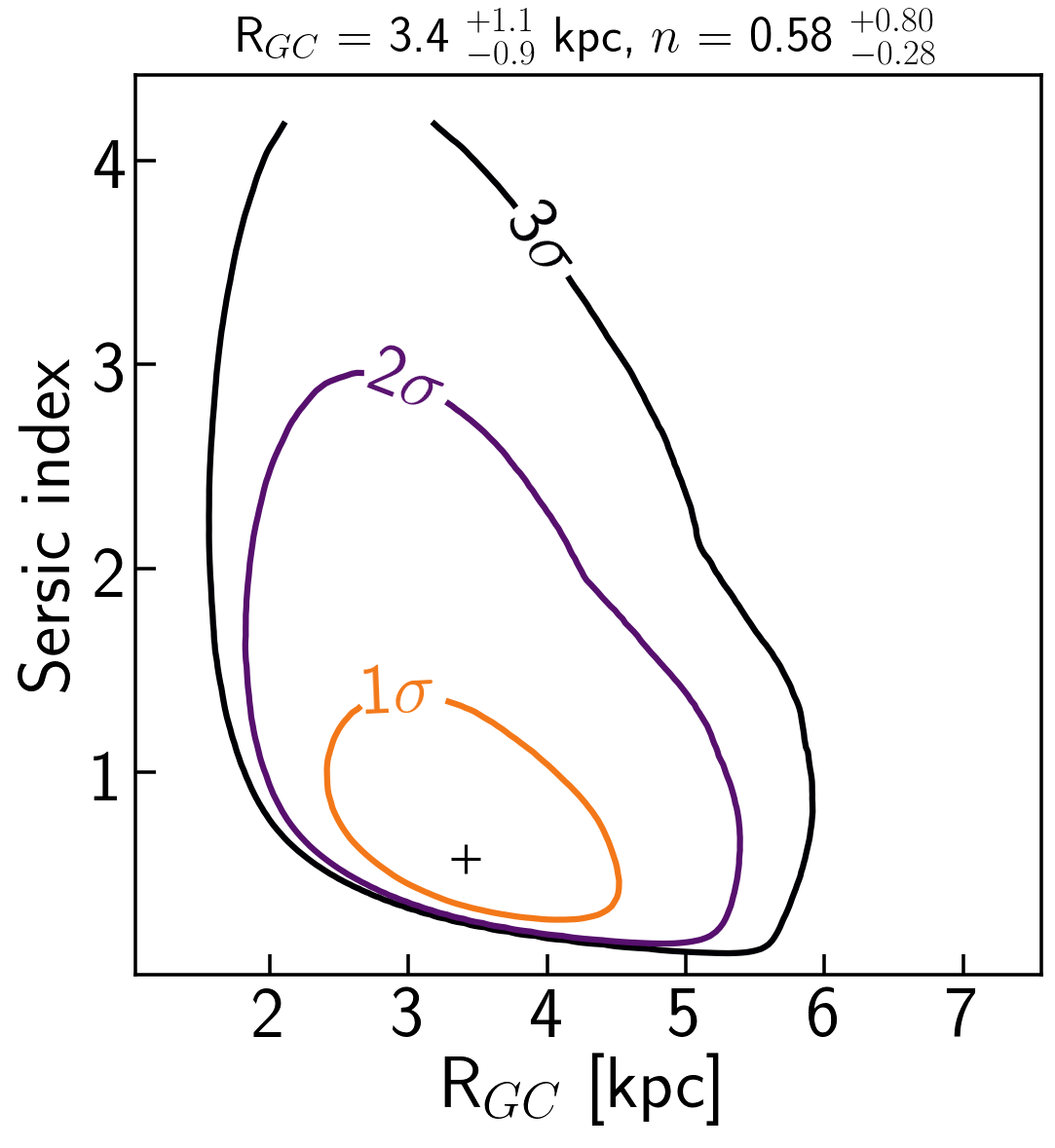}
    \end{tabular}
    \caption{Likelihood maps showing contours of constant  confidence level. The maps shown correspond to two different samples S1 (\textit{top}) and S2 (\textit{bottom}) and two different axis ratio for the spatial distribution of the GCs q=1  (circular distribution, \textit{left}) and q=0.66 (same axis ratio as the light of the galaxy, \textit{right}). The labels indicate the most likely values for the R$_{GC}$ and $n$ parameters of the GC distributions together with their 1$\sigma$ uncertainty.}
    \label{figlikelihood}
\end{figure*}

To further analyse the spatial distribution of the compact sources around DF44, we have explored the half-number radius R$_{GC}$ (i.e. the radius that contain half of the GC candidates around DF4) and the S\'ersic index of the distributions. The low number (just a few dozens) of GC candidates suggests the need to use a likelihood estimation for these quantities. The likelihood equations for a S\'ersic distribution are provided in the Appendix. Interestingly, for the R$_{GC}$ parameter, the maximum likelihood value can be estimated analytically following a simple prescription (see Eq. \ref{rgcequation}). We calculate the likelihood exploring a range for the parameters of 1 $<$ R$_{GC}$ $<$ 7 kpc and 0.05 $<$ n $<$ 4.2. We have calculated the likelihood maps for these two quantities for both sample S1 (the preferred one in this paper) and S2 (that created following vD17). In addition, we have assumed two different axis ratio distributions for the GC candidates. One assumes the GCs follow a circular distribution (i.e. q=1) while in the second we model a distribution with the same position angle and axis ratio as the underlying host galaxy (i.e. q=0.66). The contour maps are shown in Fig. \ref{figlikelihood}.

In order to produce the likelihood maps we first estimated the background contamination corresponding to samples S1 and S2. Backgrounds for S1 and S2 were estimated at radial distances from DF44 between 5R$_e$ to 10R$_e$, and the values are 0.017 kpc$^{-2}$ and 0.034 kpc$^{-2}$ respectively. This corresponds to $\sim$5, $\sim$3, $\sim$11 and $\sim$7 contaminant objects inside a radius of 10 kpc (2.5R$_e$) for S1 (q=1 and q=0.66) and S2 (q=1 and q=0.66) respectively. We also explore apertures from 2 to 3 R$_e$ to measure the background contamination and our results do not change. We randomly remove this quantity of compact sources inside 10 kpc when estimating the likelihood maps we provide in Fig. \ref{figlikelihood}. This was done by following a Gaussian probability distribution whose peak is at the mean background values and whose sigma is given by the uncertainties in measuring the background (approximately 13\% and 9\% for S1 and S2). We conducted 4000 simulations.

\begin{figure}
\centering
\includegraphics[width=\linewidth]{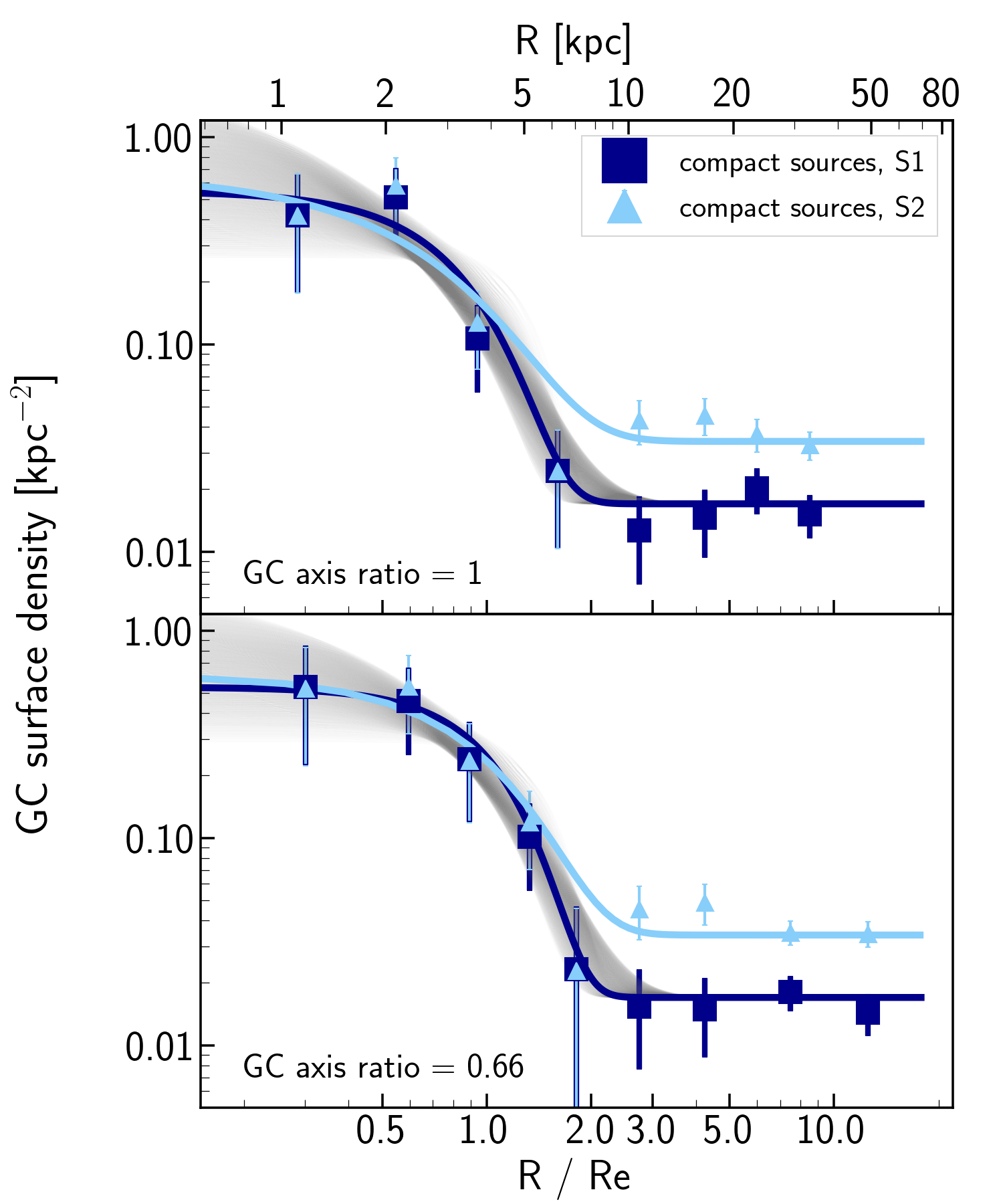}
\caption{Observed GC surface densities for the different samples and axis ratio distributions. The blue lines correspond to the S\'ersic distribution with the most likely values according to the analysis of the likelihood maps. In grey, the surface density profiles of all the S\'ersic models compatible (within 1$\sigma$) with the GCs distribution. The radial distances are shown in kpc (upper x-axis) and  normalized to the effective radius R$_e$ of the galaxy light (bottom x-axis).}
\label{radialdistributionmle}
\end{figure}

The most likely values for R$_{GC}$ and n of the GC candidates distribution, together with their uncertainties,  are provided in Table \ref{tablelikelihood}. In Fig. \ref{radialdistributionmle}, we show the GC surface density radial distribution together with the most likely solution and the rest of the solutions with 1$\sigma$ uncertainty for sample S1. The parameters describing the distribution of GCs are more affected by the assumed axis ratio distribution than by the sample used (either S1 and S2). As expected, due to the lower background contamination, the parameters on sample S1 are better determined than for S2. Interestingly, the half-number value of the GC candidates\footnote{Throughout this paper, R$_{GC}$ refers to the R$_{GC}$ measured using the S1 sample and q=0.66 (same as their host galaxy) unless otherwise explicitly stated.} (R$_{GC}$=3.1$^{+0.8}_{-0.7}$ kpc) is compatible within the error bars to the effective radius of the light distribution of DF44 (R$_e$=3.9$\pm$0.7 kpc). In other words, R$_{GC}$/R$_e$=0.8$^{+0.3}_{-0.2}$. 
 
\begin{table}
\centering
\caption {Most likely value for the S\'ersic index $n$ and half-number radius R$_{GC}$  of the GC distributions in this work and vD17. We include the parameters using different samples and/or assumptions about the GC distribution. Note that vD17 derived the GC surface density stacking two UDGs in the Coma cluster, DF44 and DFX1. Therefore, the values of the S\'ersic index and the GC half number radius quoted from that work correspond to this stacked distribution.}
\begin{tabular}{ c c c c c } 
\hline  
Reference & $q$ (axis ratio) & sample & $n$ & R$_{GC}$ [kpc] \\
\hline 
This work & 1.0 & S1 & $0.51^{+0.74}_{-0.31} $  & $2.6^{+0.8}_{-0.7}$ \\ \\
 & 1.0 & S2 & $0.72^{+1.39}_{-0.37} $  & $3.1^{+1.2}_{-1.0}$ \\ \\
 
 & 0.66 & S1 & $0.39^{+0.59}_{-0.16} $  & $3.1^{+0.8}_{-0.7}$  \\ \\
 & 0.66 & S2 & $0.58^{+0.80}_{-0.28} $  & $3.4^{+1.1}_{-0.9}$ \\
\hline
vD17 & - & - & $3.1^{+0.6}_{-0.9}$ &  $10.34^{+6.1}_{-3.3} $ \\ \\
 & - & - & $1.0$ (fixed) & $6.58^{+0.94}_{-0.94} $ \\
\hline
\end{tabular}
\label{tablelikelihood}
\end{table}

\begin{figure*}
\centering
\includegraphics[trim=70 70 70 70,clip,width=\columnwidth]{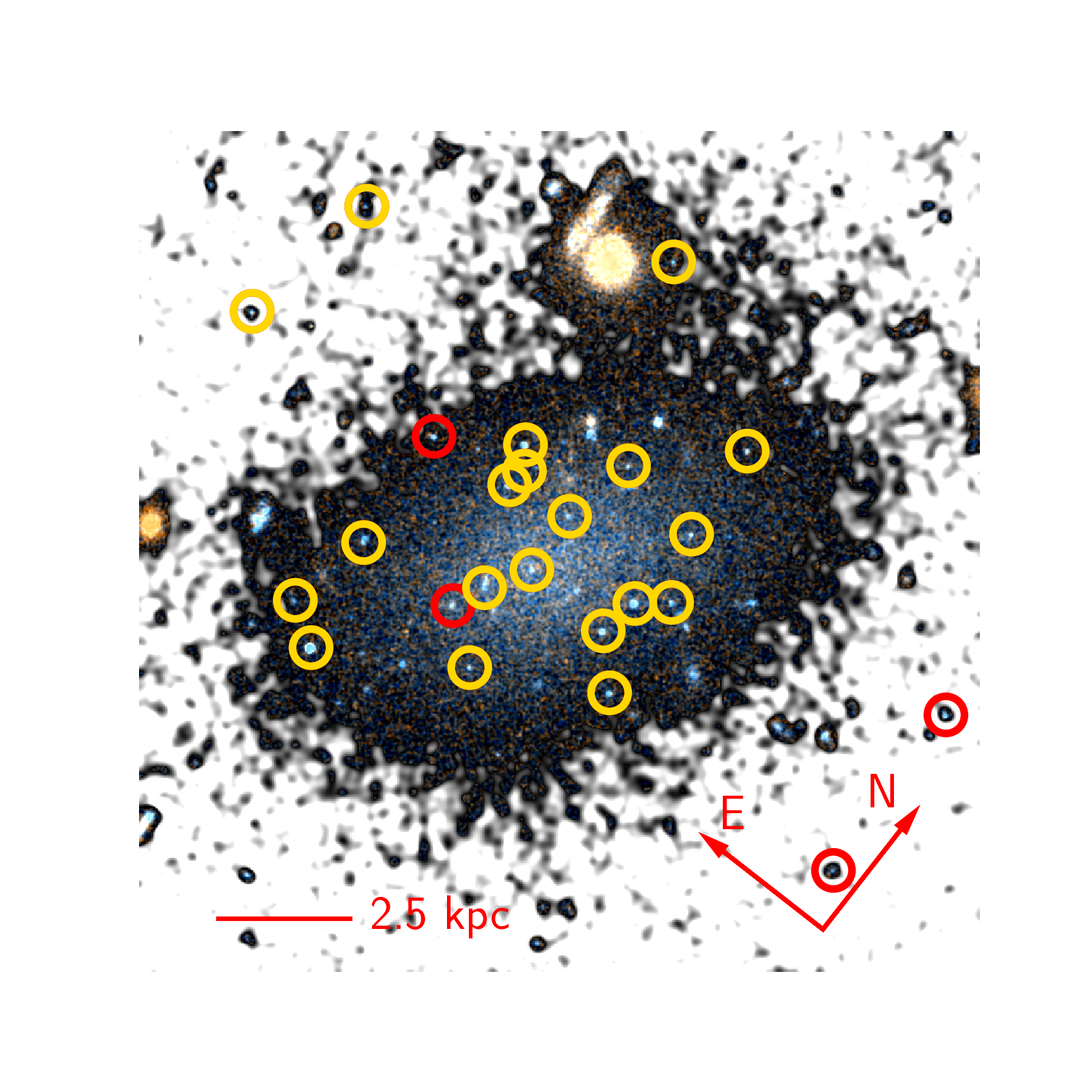}
\includegraphics[trim=70 70 70 70,clip,width=\columnwidth]{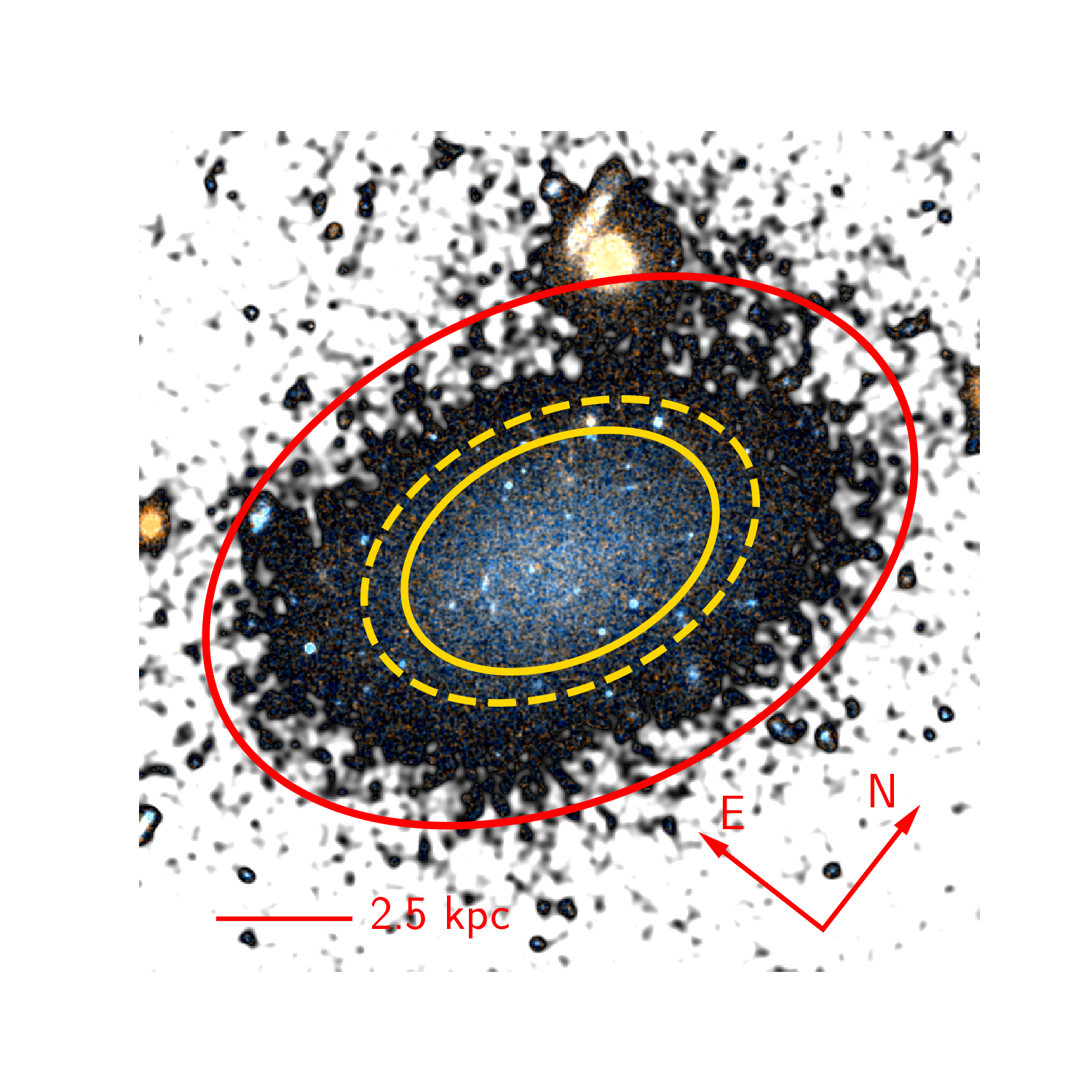}
\caption{\textit{Left}: The galaxy DF44 together with the GC candidates that compose the sample S1 (yellow circles) and the extra globular clusters added in sample S2 (red circles). All these are the GC candidates that have magnitudes brighter than \textsl{V$_{606}$}=28.0 mag and lie within a circle of radius 10 kpc. \textit{Right}: The galaxy effective radius R$_e$ (yellow dashed line) and the GCs half-number radius R$_{GC}$ (yellow solid line) estimated in this work. The red ellipse represents the GC half-number radius used in vD17. The significant difference in the values of  R$_{GC}$ between both works results in very different N$_{GC}$.}
\label{gcdistribution}
\end{figure*}

In the left panel of Fig. \ref{gcdistribution} we plot the GC candidates of sample S1 (yellow circles) and the extra GC candidates that are contained in sample S2 (following the vD17 selection criteria). Except for a few GC candidates, the remaining objects are well within the light distribution of DF44 and follow a similar distribution. In the right panel of Fig. \ref{gcdistribution} we plot the location of the ellipse containing the half-number of GCs according to our analysis for sample S1 (yellow solid line) and DF44 half light ellipse  (yellow dashed line). We also plot the ellipse containing the half-number of GCs according to vD17 (red solid line). The half-number radius R$_{GC,vD17}$=1.5R$_{e,vD17}$=7.05 kpc that was used by vD17 is not supported by the distribution of the GC candidates we detect. As we will show later on in the text, it is this significant difference in the estimation of R$_{GC}$ between vD17 and the present work which leads to a strong disagreement in the total number of GCs found by vD17 and ourselves.

\subsection{Globular cluster luminosity function}

With the GC candidates selected and their spatial distribution characterised, the next step is to calculate the total number of GCs (N$_{GC}$) around DF44.  To compute N$_{GC}$ we require knowledge of the GC luminosity function (GCLF) in DF44. This function can be approximated by a Gaussian distribution \citep{1993AJ....105.1358S}. The area of the distribution represents the total number of GCs. The peak of the distribution can be used as a standard candle and is only weakly dependent on host galaxy mass \citep[e.g.][]{gclf1,rejkuba2012}. This peak is located at M$_V$=-7.5 mag (Johnson V filter and in the Vega system), which corresponds to m$_V$ = 27.5 mag at the distance of the Coma Cluster \citep{peng2011}.
 
In order to transform the previous numbers from \textsl{V}(Vega) to \textsl{V$_{606}$}(AB), we use the following assumptions. First, we assume that the GCs of DF44 are old ($\sim$12 Gyr) and well described by simple stellar population models. From MILES  models \citep{vazdekis2016}, we get (\textsl{V}-\textsl{I})(Vega)$\sim$0.9 mag for a wide range of metallicities (-2$<$Z$<$-0.5). Next, we use the transformation provided by \citet{harris2018} to get the magnitude of the peak of the GCLF in   \textsl{V$_{606}$}(Vega):

\begin{equation}
    V_{606}(Vega) = V - 0.263 \times (V - I) + 0.091 
\end{equation}

Finally, we transform that value to the AB system by using the handbook of the WFC3 instrument \citep{wfc3}:

\begin{equation}
V_{606}(AB) = V_{606}(Vega) + 0.093
\end{equation}

\begin{figure*}
    \centering
    \begin{tabular}{l l}
        \textbf{i. S1 sample, GC axis ratio = 1} & \textbf{ii. S1 sample, GC axis ratio = 0.66} \\\\
        \includegraphics[width=0.4\linewidth]{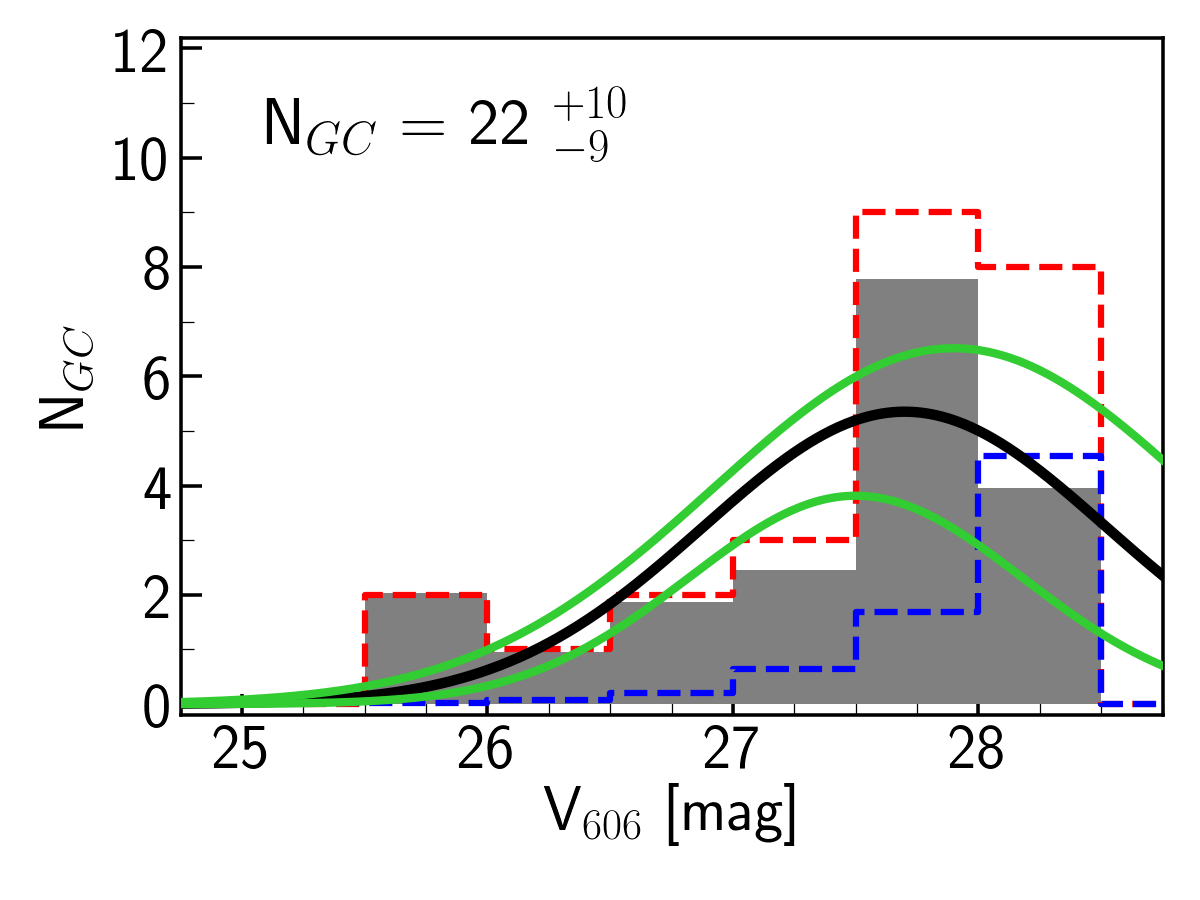} & \includegraphics[width=0.4\linewidth]{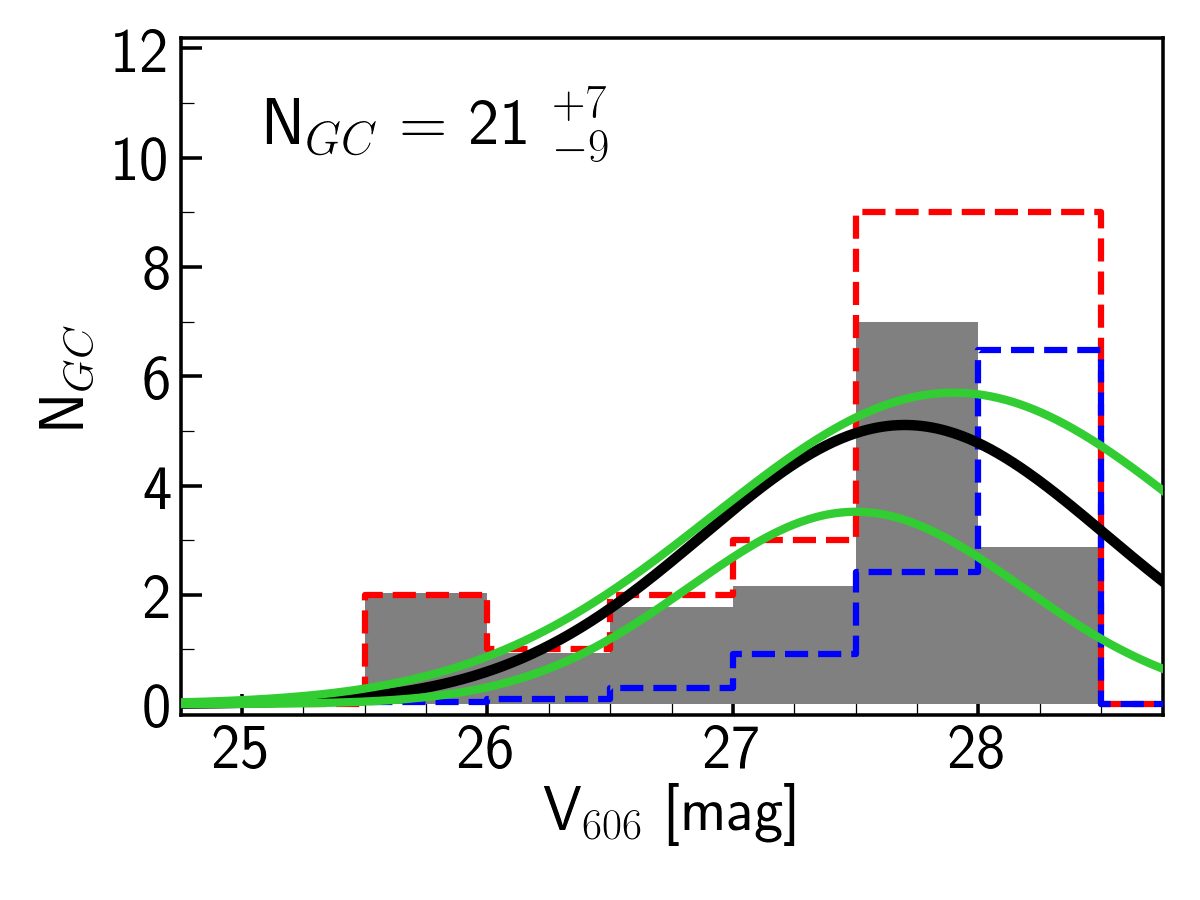} \\\\\\
        \textbf{iii. S2 sample, GC axis ratio = 1} & \textbf{iv. S2 sample, GC axis ratio = 0.66} \\\\
        \includegraphics[width=0.4\linewidth]{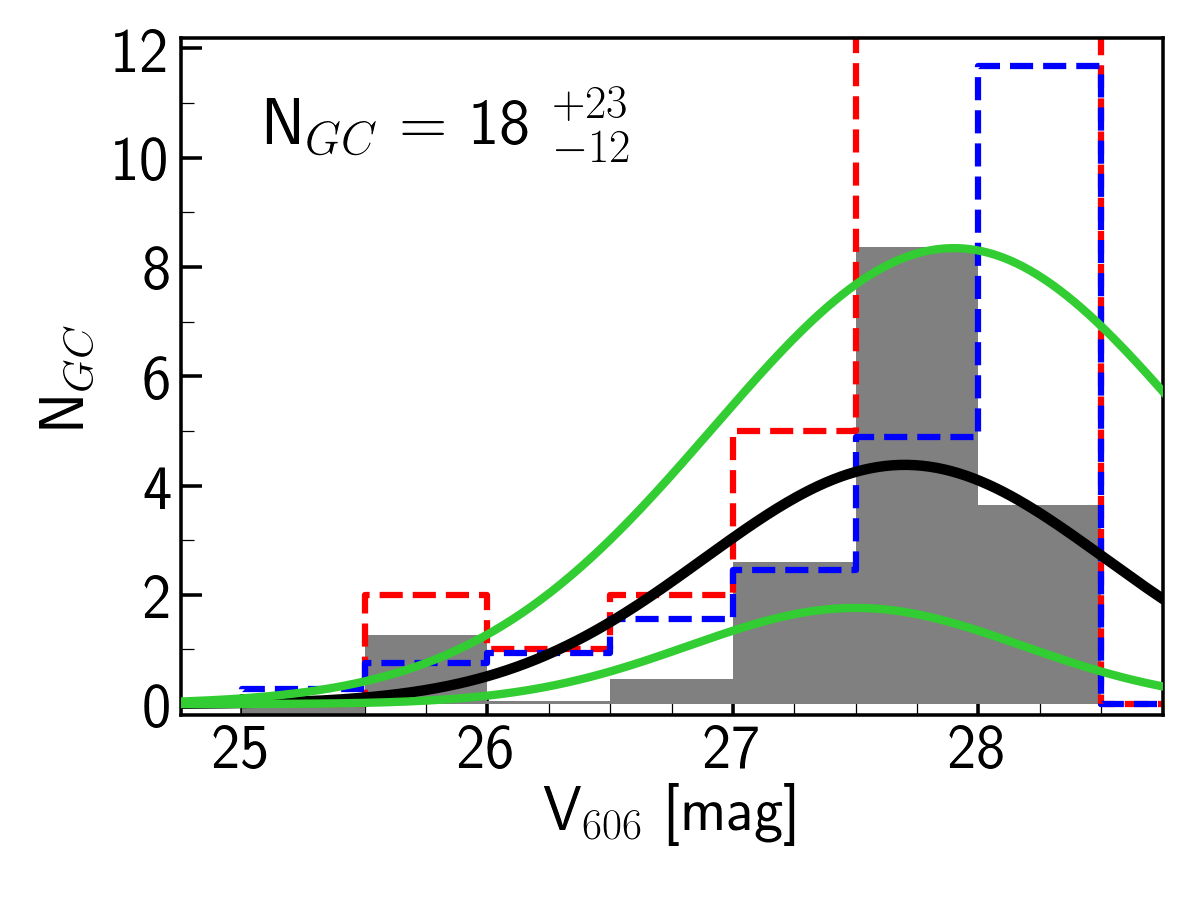} & \includegraphics[width=0.4\linewidth]{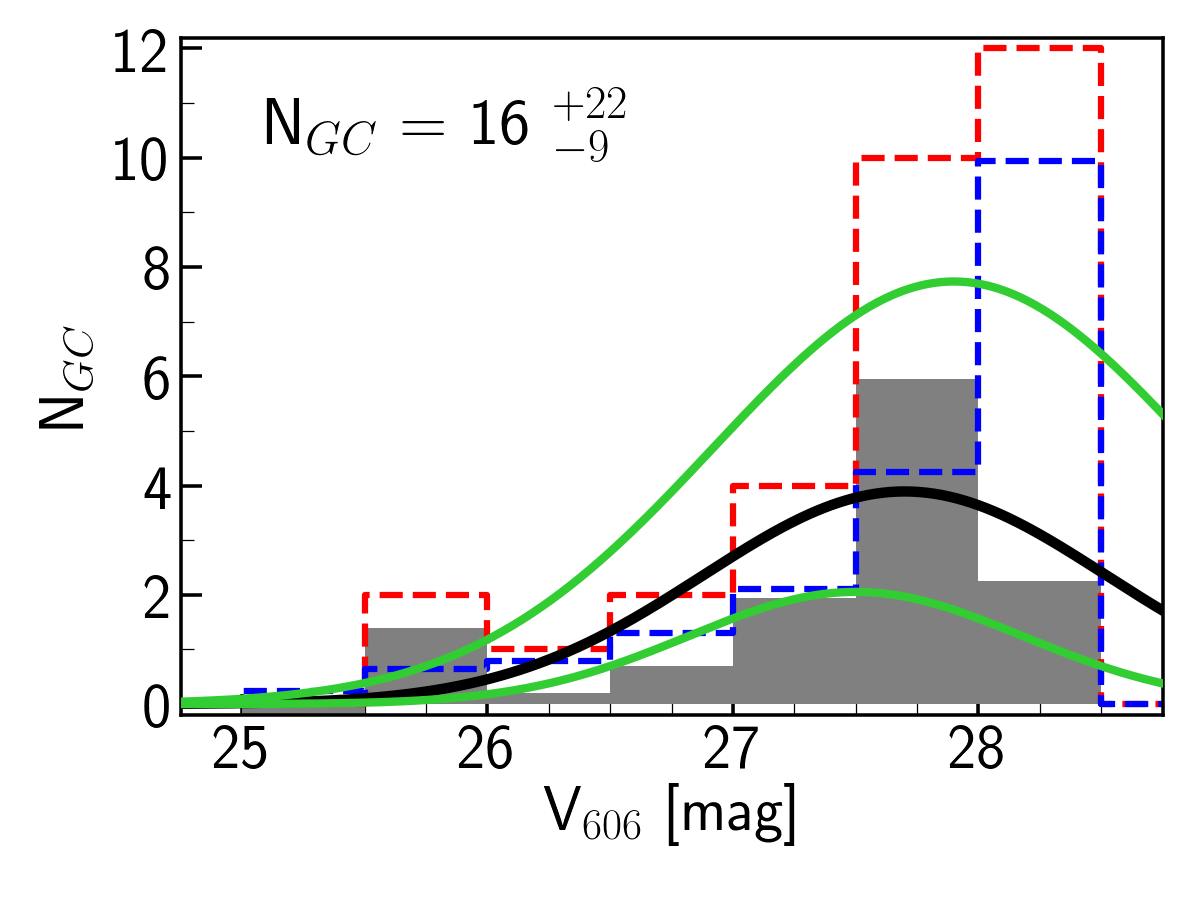}
    \end{tabular}
    \caption{Globular cluster luminosity function (GCLF) for the S1 and S2 samples with different GC axis-ratios. In red, the observed distribution of compact sources. The blue dashed line corresponds to the background sources while in grey we show the background corrected distribution in the number of compact sources.  The black solid line indicates the GCLF corresponding to the calculated average number of the GCs while the green lines enclose the uncertainty in deriving N$_{GC}$ (see text for details).}
    \label{lfunctiongc}
\end{figure*}

This give us a magnitude for the peak of the GCLF in the AB system of <\textsl{V$_{606}$}>$\sim$27.45 mag. In order to estimate N$_{GC}$, we consider only the GC candidates inside R$<$3R$_{GC}$. We provide in Table \ref{gccandidates} the catalogue of GC candidates for S1 and q=0.66. We use a radial distance of  3R$_{GC}$ as this radius encloses (for the S\'ersic index value n$\lesssim$0.7 describing the compact sources distribution around DF44) 99\% of the expected objects \citep{2001MNRAS.326..869T}. We plot the derived luminosity functions for different cases (S1 and S2 samples with q=1 and q=0.66) in Fig. \ref{lfunctiongc}. Together with the observed luminosity function (red dashed line), we show the background contribution to this function (blue dashed line). Background contamination for each luminosity function was estimated using the identified compact sources further than the radius 3R$_{GC}$ from DF44 in the image. This allows us to have a statistical significant correction of the background, particularly for the most luminous sources. The observed luminosity functions corrected by the background is shown in grey in Fig. \ref{lfunctiongc}. This corrected luminosity function has also been corrected for incompleteness down to magnitude \textsl{V$_{606}$}=28.5 mag.

\begin{table*}
\centering
\caption{Catalogue of GC  candidates in S1 sample and q=0.66. The columns represent R.A. (a), Declination (b), distance to DF44 center as defined in the previous coordinates (c), \textsl{V$_{606}$} magnitude (d) and its error (e), compactness parameter (f) and its error (g), ellipticity (h), \textsl{V$_{606}$}-\textsl{I$_{814}$} colour (i) and its error (j). Those GC candidates without a colour estimation correspond to objects that were not detected in the \textsl{I$_{814}$} image.} 
\begin{tabular}{ c c c c c c c c c c  } 
\hline  
RA & DEC & R & \textsl{V$_{606}$} & $e\_\textsl{V$_{606}$}$ & $\Delta m_{4-8}$ & $e\_\Delta m_{4-8}$  & ell & \textsl{V$_{606}$}-\textsl{I$_{814}$} & $e\_\textsl{V$_{606}$}-\textsl{I$_{814}$}$\\
$h$ $m$ $s$ & $d$ $m$ $s$ & \arcsec & mag & mag & mag & mag & - & mag & mag \\
(a) & (b) & (c) & (d) & (e) & (f) & (g) & (h) & (i) & (j) \\
\hline
13h 00m 58.33s  & +26d 58m 25.83s  & 10.32 & 23.92 & 0.002 & 0.38 & 0.004 & 0.03 & 0.33 & 0.01 \\
13h 00m 57.70s  & +26d 58m 34.80s  & 3.882 & 25.74 & 0.013 & 0.43 & 0.023 & 0.02 & 0.41 & 0.03 \\ 
13h 00m 58.22s  & +26d 58m 36.85s  & 4.596 & 25.98 & 0.017 & 0.36 & 0.029 & 0.05 & 0.44 & 0.04 \\
13h 00m 58.18s  & +26d 58m 35.30s  & 3.482 & 26.49 & 0.027 & 0.46 & 0.045 & 0.07 & 0.39 & 0.08\\
13h 00m 57.72s  & +26d 58m 33.23s  & 3.787 & 26.71 & 0.033 & 0.42 & 0.056 & 0.03 & 0.41 & 0.08\\
13h 00m 57.59s  & +26d 58m 31.58s  & 6.165 & 26.95 & 0.042 & 0.54 & 0.066 & 0.09 & 0.19 & 0.13\\
13h 00m 57.62s  & +26d 58m 35.70s  & 5.232 & 27.27 & 0.056 & 0.52 & 0.089 & 0.18 & 0.21 & 0.17\\
13h 00m 57.72s  & +26d 58m 41.95s  & 8.249 & 27.39 & 0.063 & 0.48 & 0.102 & 0.15 & 0.47 & 0.17\\
13h 00m 58.40s  & +26d 58m 30.18s  & 7.932 & 27.47 & 0.067 & 0.45 & 0.111 & 0.29 & 0.21 & 0.21 \\
13h 00m 57.98s  & +26d 58m 33.36s  & 1.202 & 27.55 & 0.073 & 0.47 & 0.119 & 0.14 & 0.55 & 0.17 \\
13h 00m 57.69s  & +26d 58m 38.17s  & 5.369 & 27.57 & 0.074 & 0.46 & 0.122 & 0.04 & 0.81 & 0.15 \\
13h 00m 58.18s  & +26d 58m 36.07s  & 3.659 & 27.58 & 0.074 & 0.14 & 0.149 & 0.34 & --- & --- \\
13h 00m 58.22s  & +26d 58m 45.75s  & 11.91 & 27.67 & 0.080 & 0.48 & 0.131 & 0.34 & --- & --- \\
13h 00m 58.05s  & +26d 58m 31.70s  & 3.152 & 27.82 & 0.093 & 0.58 & 0.143 & 0.11 & --- & --- \\
13h 00m 57.95s  & +26d 58m 38.69s  & 4.190 & 27.85 & 0.096 & 0.37 & 0.166 & 0.09 & --- & --- \\
13h 00m 57.94s  & +26d 58m 29.01s  & 5.483 & 27.89 & 0.099 & 0.47 & 0.163 & 0.31 & 0.59 & 0.23 \\
13h 00m 58.45s  & +26d 58m 26.84s  & 10.64 & 27.89 & 0.099 & 0.57 & 0.154 & 0.40 & --- & --- \\
13h 00m 58.99s  & +26d 58m 40.10s  & 16.50 & 27.94 & 0.104 & 0.58 & 0.161 & 0.04 & --- & --- \\
13h 00m 58.00s  & +26d 58m 35.82s  & 1.452 & 27.95 & 0.104 & 0.56 & 0.163 & 0.25 & --- & --- \\
13h 00m 57.95s  & +26d 58m 46.26s  & 11.76 & 28.00 & 0.110 & 0.56 & 0.171 & 0.07 & --- & --- \\
13h 00m 57.73s  & +26d 58m 27.15s  & 8.092 & 28.14 & 0.124 & 0.21 & 0.240 & 0.30 & --- & --- \\
13h 00m 58.17s  & +26d 58m 24.90s  & 10.11 & 28.14 & 0.125 & 0.66 & 0.184 & 0.16 & --- & --- \\
13h 00m 58.20s  & +26d 58m 26.91s  & 8.453 & 28.16 & 0.127 & 0.53 & 0.201 & 0.22 & --- & --- \\
13h 00m 58.37s  & +26d 58m 19.74s  & 16.04 & 28.20 & 0.133 & 0.66 & 0.196 & 0.21 & --- & --- \\
13h 00m 58.14s  & +26d 58m 37.86s  & 4.337 & 28.27 & 0.142 & 0.15 & 0.284 & 0.19 & --- & --- \\
13h 00m 58.90s  & +26d 58m 35.30s  & 14.15 & 28.30 & 0.145 & 0.31 & 0.260 & 0.15 & --- & --- \\
13h 00m 57.60s  & +26d 58m 39.04s  & 6.948 & 28.39 & 0.157 & 0.26 & 0.292 & 0.26 & --- & --- \\
13h 00m 58.16s  & +26d 58m 34.82s  & 3.159 & 28.42 & 0.161 & 0.64 & 0.241 & 0.36 & --- & --- \\
\hline
\end{tabular}
\label{gccandidates}
\end{table*}

Once we have the GCLF corrected for background and incompleteness, we count all the GCs up to the location of the peak <\textsl{V$_{606}$}>. This number is then multiplied by 2 to get the total number of GCs (N$_{GC}$). For the location of the peak and the width of the GC distribution ($\sigma_{GC}$), we use the numbers given by vD17 (i.e.  <\textsl{V$_{606}$}>=27.7$_{-0.2}^{+0.2} $ mag and $\sigma_{GC}$=0.82$_{-0.15}^{+0.16}$). The total number of GCs is then all compact sources (background and incompleteness corrected) within R$<$3R$_{GC}$ which are within the \textsl{V$_{606}$} magnitude range \textsl{V$_{606}$}-3$\sigma_{GC}$$<$\textsl{V$_{606}$}$<$ <\textsl{V$_{606}$}> multiplied by 2. To estimate the uncertainties on N$_{GC}$ we take into account the uncertainties on the number of observed compact sources within 3R$_{GC}$ and in the background, the uncertainty in R$_{GC}$ we have estimated above and also the uncertainties in <\textsl{V$_{606}$}> and $\sigma_{GC}$ of the GCLF.

In Fig. \ref{lfunctiongc} we show the Gaussian distributions corresponding to the number of GCs derived for each sample (black solid line) together with the Gaussian distributions compatible with the minimum and maximum number of GCs (within 1$\sigma$) compatible with the data (green lines). For all our samples and spatial configurations the total number of GCs is very modest. In particular, for our preferred sample S1 and spatial configuration q=0.66, N$_{GC}$=21$^{+7}_{-9}$. This number is in stark contrast with the value reported by vD17: 74$\pm$18. We will expand on this discrepancy in section \ref{thisworkvsvd17}.

\subsection{The average colour of the population of GCs around DF44}

\begin{figure}
\centering
\includegraphics[width=\linewidth]{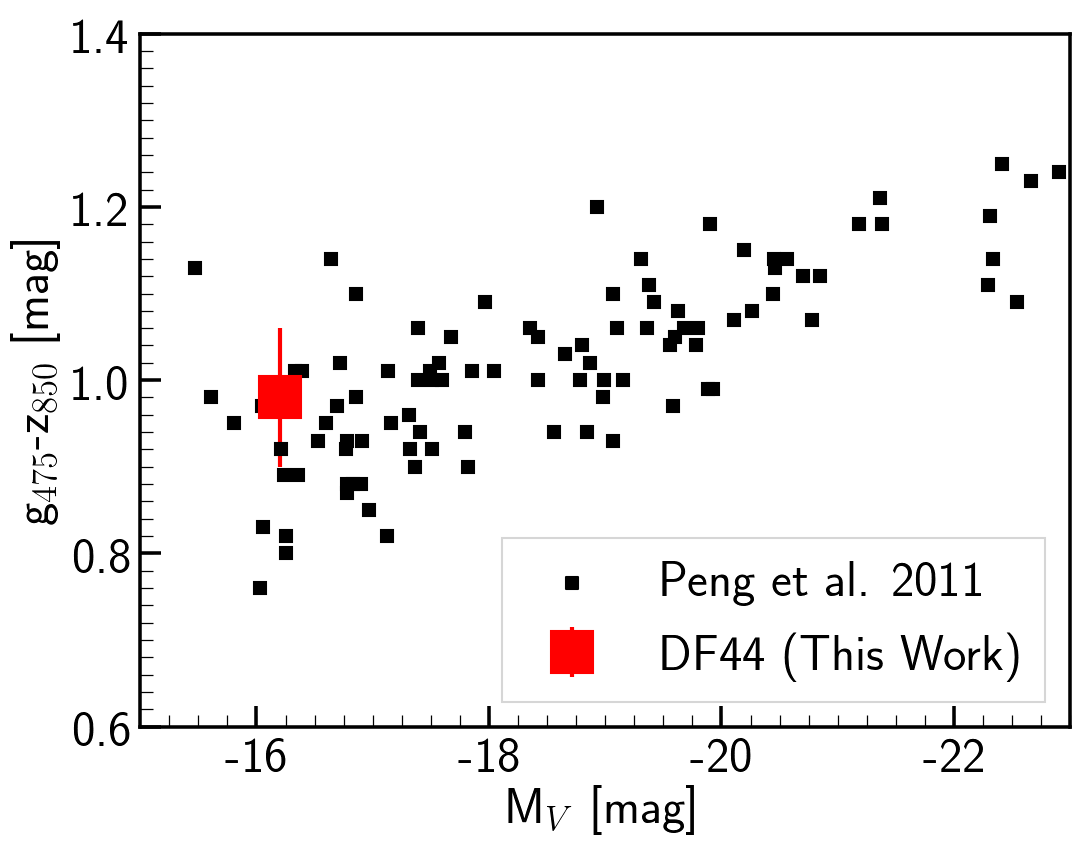}
\caption{The average colour of GCs in DF44 (red square) compared with the colours of the GCs in Coma cluster galaxies \citep{peng2011}. The average colour of the DF44 GCs follows the observed trend for GCs in Coma galaxies, having a colour compatible with GCs of galaxies of the same luminosity.}
\label{gccolors}
\end{figure}

Another interesting exercise we can conduct on the GC population around DF44 is to estimate its average colour and compare it with other GC samples. Fig. \ref{gccolors} shows the average colour \textsl{g$_{475}$}-\textsl{z$_{850}$} of the GC population of DF44 compared to other Coma Cluster galaxies \citep{peng2011}. There is a clear trend between the average colour of the GCs and the total luminosity of their host galaxies. 

To estimate the average colour \textsl{g$_{475}$}-\textsl{z$_{850}$} of the GCs of DF44, we start using the observed colour  \textsl{V$_{606}$}-\textsl{I$_{814}$} = 0.40$\pm$0.04 mag. This value is obtained using the brightest (i.e. least affected by the background) GC candidates. We only used GCs with \textsl{V$_{606}$}$<$27.0 mag. This colour is also consistent with that given in vD17. The observed colour  \textsl{V$_{606}$}-\textsl{I$_{814}$} was transformed to \textsl{g$_{475}$}-\textsl{z$_{850}$} using  the following transformations: (i) \textsl{g$_{475}$}-\textsl{I$_{814}$}=\textsl{V$_{606}$}-\textsl{I$_{814}$}$\times$1.852+0.096 \citep{blak2010} and (ii) \textsl{g$_{475}$}-\textsl{z$_{850}$}=\textsl{g$_{475}$}-\textsl{I$_{814}$}$\times$1.023+0.128 \citep{beasley2016b}. Consequently, we end up having an average colour \textsl{g$_{475}$}-\textsl{z$_{814}$}=0.98$\pm$0.08 for the GCs around DF44. This colour corresponds to a metallicity of [M/H]$\sim$-0.9, using the photometric predictions of the MILES stellar population models \citep{vazdekis2016}, for an old (12 Gyr) SSP. Fig. \ref{gccolors} shows that the average colour of the DF44 GCs are in perfect agreement with other galaxies in the Coma Cluster of similar luminosity.
 
\section{Discussion}

\subsection{The halo mass of DF44 based on the GC population}

We can use the total number of GCs to estimate the halo mass of DF44,
considering that the total stellar mass contained in the GCs in a galaxy
scales linearly with galaxy halo mass \citep[see e.g.][]{harris2017}. This observation has been shown to be particularly useful
for inferring the halo masses of UDGs \citep{beasley2016}. According to \citet{harris2017}, the relation between both masses is linked through the following equation: M$_{GCs}$/M$_{halo}$=3.9$\times$10$^{-5}$. With the total number of GCs (N$_{GC}$=21$^{+7}_{-9}$) derived here, we have used the following approximation. We assume a GC mean mass $<M_{GC}>$=2$\times$10$^5$ M$_{\odot}$ \citep{jordan2007}, which is multiplied by N$_{GC}$ to get M$_{GCs}$. This gives a total halo mass of M$_{halo}$=1.1$^{+0.4}_{-0.5}$$\times$10$^{11}$M$_{\odot}$.

This estimation for the mass of the halo of DF44 is in agreement (within the error bars) with that derived from the velocity dispersion of the stars of the galaxy (see Table \ref{df44mass}), and removes the tension between the kinematic mass and the mass derived from the large number of DF44 GCs identified by vD17. Recently, a dwarf-like dark matter halo ($\sim$ 10$^{11}$M$_{\odot}$) has been confirmed by \citet{akos}.

\begin{table*}
\centering
\caption{Summary of the DF44 halo mass estimates (M$_{halo}$) using different mass proxies from vD16, vD17, vD19 and this work. S$_{N}$ indicates the specific frequency of the GC system in the \textsl{V} band. vD19 noted that motivated by the new measurements of the velocity dispersion of the galaxy DF44, they uncovered an error in vD16 and the corrected velocity dispersion from the old data is $\sigma=42^{+7}_{-7}$ km~s$^{-1}$.}
\begin{tabular}{ c c c c c c } \hline  
Ref & N$_{GC}$ & S$_{N}$ & $\sigma$ & M$_{halo}$ & Mass proxy \\
& & & (km s$^{-1}$) & (M$_{\odot}$) & \\
\hline 
vD16 & 94$^{+25}_{-20}$ & 31.1$^{+8.3}_{-6.6}$ & 47$^{+8}_{-6}$ & 1.0$\times$10$^{12}$ & kinematics \\ \\
vD17 & 74$^{+18}_{-18}$  & 24.5$^{+6.0}_{-6.0}$ & - & 5.0$\times$10$^{11}$ & GC number count\\ \\
vD19 & - & - & 33$^{+3}_{-3}$ & 1.6 $^{+5.0}_{-1.2}$$\times$10$^{11}$ & kinematics \\ \\
This work & 21$^{+7}_{-9}$  & 7.0$^{+2.3}_{-3.0}$ & - & 1.1$^{+0.4}_{-0.5}$$\times$10$^{11}$ & GC number count\\ 
\hline 
\end{tabular}
\label{df44mass}
\end{table*}

\subsection{DF44 GCs in comparison with other GC systems}

\begin{figure*}
\centering
    \includegraphics[trim=0 20 50 20,clip,clip,width=0.8\linewidth]{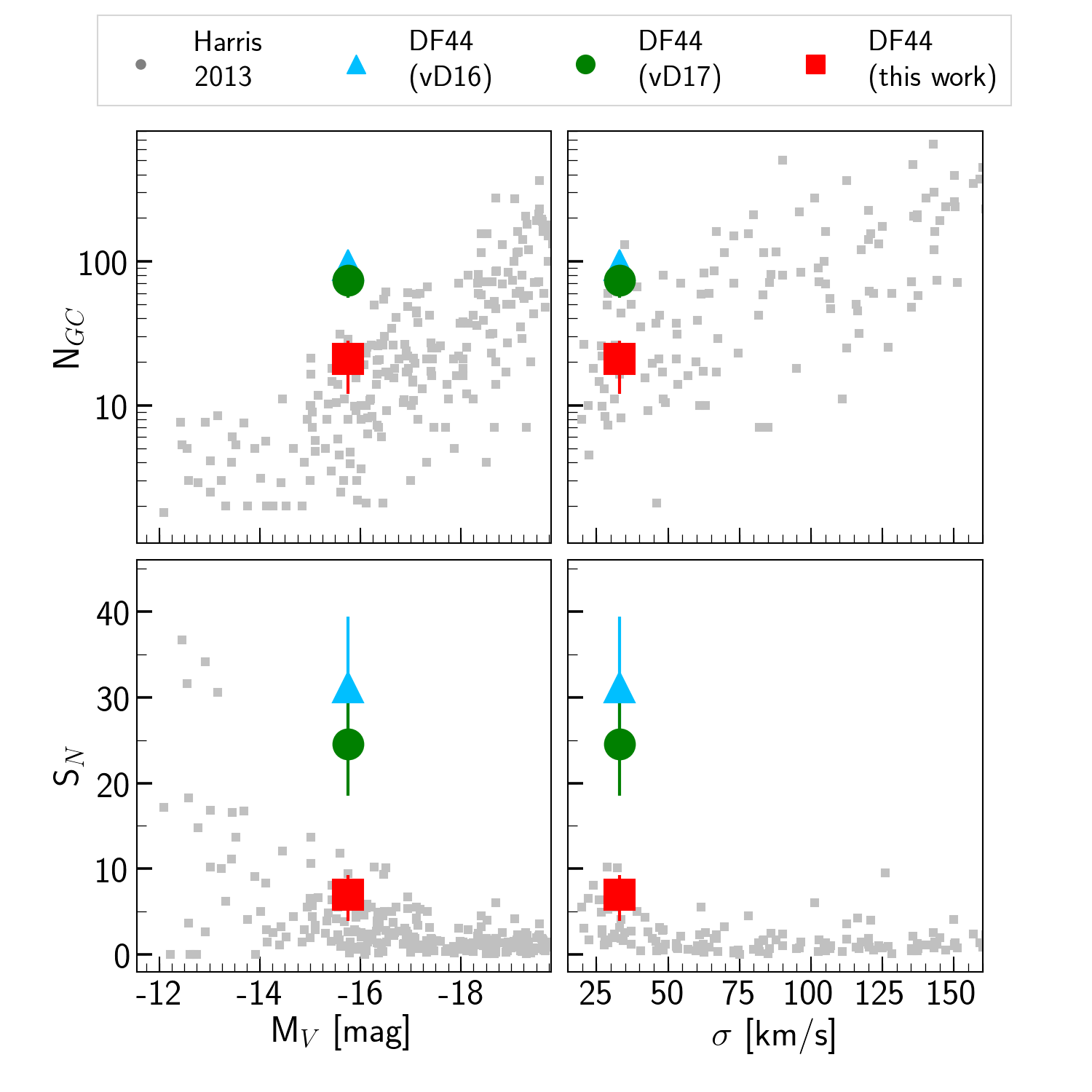}
    \caption{Total number of GCs (N$_{GC}$) and GCs specific luminosity frequency in the \textsl{V}-band ($S_N$)  for galaxies from \citet{harris2013} (grey points). DF44 is also shown as a red square (this work), green circle (vD17) and  blue triangle (vD16). DF44 absolute magnitude (M$_V$=-16.2 mag) and velocity dispersion ($\sigma$=33 km s$^{-1}$) are from vD17 and vD19 respectively. Based on vD17, the total number of GCs of DF44 is higher than the average number for galaxies with similar velocity dispersion (and therefore mass) and luminosity. In this work, we used the same data-set as vD17 and find N$_{GC}$=21$^{+7}_{-9}$ for DF44. With this new estimation, the N$_{GC}$ of DF44 is consistent with those shown on galaxies with similar luminosities and velocity dispersion. }
    \label{df44incontext}
\end{figure*}

As we reported in Sect. 3.5, the average colour of the GC population of DF44 is in excellent agreement with the expected colour for galaxies with similar luminosities. Another aspect to address is whether both the number of GCs and their specific frequencies fit well with other galaxies with similar characteristics. In Fig. \ref{df44incontext} we explore this. We have used the compilation by \citet{harris2013} which gives the number of GCs and their specific frequency against the luminosity of the galaxies in the \textsl{V}-band along with their central velocity dispersions. As stated before, we use for the total luminosity of DF44 (M$_V$=-16.2 mag) and its velocity dispersion ($\sigma$=33$^{+3}_{-3}$ km s$^{-1}$), the values reported by vD17 and vD19 respectively. Using these values, and with the number of GCs measured in this work, DF44 is within the observed the relations found for other galaxies.

\begin{figure*}
  \includegraphics[trim=0 20 50 20,clip,clip,width=0.8\linewidth]{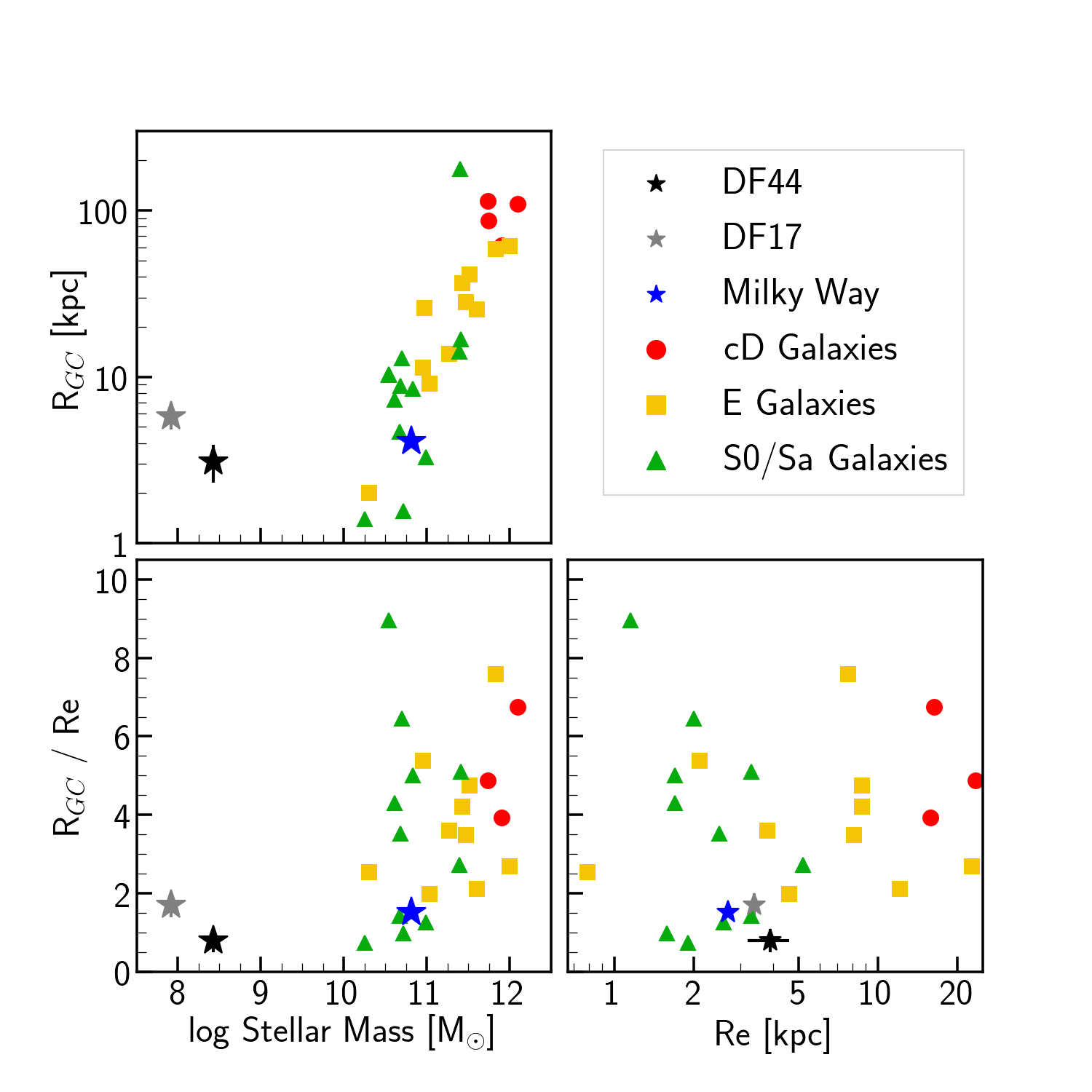}
  \caption{GC distribution around different types of galaxies with different effective radii and stellar masses. We show a compilation  presented by \citet{forbes2017}, DF44 (this work) and DF17 (\citealp{peng2016}). It can be seen that DF44 shows R$_e$/R$_{GC}$$\sim$1. Values similar to this are observed in other nearby galaxies, such as two S0 galaxies: NGC7332 and NGC5473.}
  \label{gcratio}
\end{figure*}

Another property we can compare with other galaxies is the ratio between the half-number radius of GCs for DF44 compared to the effective radius R$_e$ of the host galaxy. We find that  R$_{GC}$/R$_e$=0.8$^{+0.3}_{-0.2}$. This ratio is compared with other galaxies in Fig. \ref{gcratio} using the sample given in \citet{forbes2017}. Unfortunately, this kind of work has not been conducted for galaxies with similar stellar masses to DF44 \citep[with exception of DF17; ][]{peng2016}. Both UDGs share the property that  R$_{GC}$/R$_e$$<$2, something which is only observed for some nearby relatively massive galaxies, including our own Milky Way. It remains as an interesting exercise to fill the gap between the massive galaxies and the dwarf population in terms of the ratio between both radii to explore whether UDGs have an anomalously low spatial extension for their GCs or not.

\subsection{Different estimates for the radial distribution of GCs lead to much smaller N$_{GC}$ than in vD17}
\label{thisworkvsvd17}

The main result of this paper is that the number of GCs around DF44, N$_{GC}$=21$^{+7}_{-9}$, is compatible with the number of GCs expected for galaxies with the luminosity and velocity dispersion of DF44 (see previous section). This results is in disagreement with the high value of GCs found by vD17 (N$_{GC}$=74$^{+18}_{-18}$). What is the origin of this discrepancy?

We have explored whether the selection of GC candidates around DF44 based on different compactness criteria can explain the different values of N$_{GC}$ obtained by us and vD17. To do this, we estimated N$_{GC}$ using our sample S2 which uses the same selection criteria as vD17. We get N$_{GC}$=18$^{+23}_{-12}$ (assuming a projected circular distribution for the GCs; q=1) and N$_{GC}$=16$^{+22}_{-9}$ (assuming the GCs have the same axial distribution as DF44 itself; q=0.66). As expected, the larger background contamination in sample S2 increases the uncertainty in measuring N$_{GC}$ compared to sample S1. However, even considering the uncertainties, the maximum number of globular clusters is not larger than $\sim$40. This is around a factor of 2 smaller than the value reported by vD17. 

vD17 indicate in their paper that their total number of GCs in DF44 is four times the number of observed GCs (contamination-corrected) within R$<$1.5R$_{e,vD17}$ and \textsl{V$_{606}$}$<$27.6 mag (i.e. N$_{GC,obs}$=18.5). It is interesting to note that this number is already larger (around a factor of 2) than the values we find for the number of observed GCs to a similar radial distance and magnitude than vD17 (see e.g. our grey histograms in Fig. \ref{lfunctiongc}). This discrepancy is potentially significantly affected by the different background contamination corrections both works have applied. 
Unfortunately, we can not perform a direct comparison between our GC detections and vD17 since these data are not provided in vD17.

The rational for why vD17 multiplied by a factor of 4 their N$_{GC,obs}$  is as follows. First, vD17 multiply by 2 in order to correct for the number of GCs missed after the peak of the GCLF. We have also done this correction to our N$_{GC}$ values reported above. However, since our N$_{GC,obs}$ is a factor of $\sim$2 less than vD17, the discrepancy remains. Second, vD17 multiply again by another factor of 2 to account for GCs which are beyond their spatial selection criterion of R=1.5R$_{e,vD17}$ (where R$_{e,vD17}$=4.7 kpc according to vD17). This spatial selection criterion, R=1.5R$_e$, corresponds to their R$_{GC}$ assuming GCs follow an exponential declining distribution from the center of the distribution. R$_{GC}$ for vD17 is 7 kpc. It is in this last correction where both works disagree fundamentally. While vD17 assumes an exponential distribution (as they do not measure the actual distribution of their GCs) we have measured this quantity. We find the GCs follow a S\'ersic distribution with a S\'ersic index (n$\sim$0.5) significantly lower than an exponential. In addition, we have also measured R$_{GC}$ directly (without assuming an exponential distribution), finding a value of 2.5-3.5 kpc. In practical terms this means that there is not a significant number of GCs beyond 7 kpc.

To summarize, according to our measurements there is no need for multiply by another factor of 2 to correct for missing GCs at large radii. Our estimation of the total number of GCs is around a factor of 4 smaller than vD17. Half of the difference is easily explained by the correction that vD17 made for the spatial distribution of GCs that here we find is not necessary. The other half is due to a different background correction of their sample.

\section{Conclusion}

The existence of UDGs with relatively large numbers of GCs, when compared to the expectations based on their luminosity or dynamical mass, has remained a puzzle.
The general properties of UDGs are consistent with them being the low surface brightness counterparts of regular dwarf galaxies. However, the existence of GC-rich UDGs has been an argument to support the idea that at least some of these galaxies could be intrinsically different from dwarfs. In this paper, we have explored this issue in detail by revisiting one of the most iconic UDGs, DF44 in the Coma cluster. DF44 has been claimed to have a dark matter halo similar in mass to that of the Milky Way, and could, therefore be a potential candidate "failed Milky Way" (vD16). More generally, such a massive halo would make DF44 an extreme outlier in stellar mass -- halo mass relations \citep{beasley2016}.

Through a detailed analysis of the GC candidates around this object, we have found a total number of GCs of only N$_{GC}$=21$^{+7}_{-9}$, which is a factor of 4 lower than previous measurements (vD17). We believe that the vD17 assumption of a large R$_{GC}$, based on the results in the literature for non-UDG galaxies, has led the authors to over-estimate N$_{GC}$. The significant reduction in the number of GCs found in our work is in good agreement with the expectation for objects with similar central velocity dispersions ($\sigma$=33 km s$^{-1}$; vD19). A smaller N$_{GC}$ resolves the strong tension with respect to previous claims about the amount of dark matter in the halos of these objects based either in their number of GCs or in their dynamical mass. In addition, we have found that the colour and specific frequency of DF44's GCs is in agreement with galaxies of similar luminosities. Based on this analysis, we conclude that DF44 is compatible with being a "regular" low surface brightness dwarf galaxy and the need for a new class of galaxies to account for this type of object is not yet required.

\section*{Acknowledgements}

We thank the referee for their detailed reading of our manuscript and for a number of excellent suggestions to improve the quality of the analysis. T.S., I.T., R.F.P and J.H.K. acknowledge financial support from the European Union's Horizon 2020 research and innovation programme under Marie Sk\l odowska-Curie grant agreement No 721463 to the SUNDIAL ITN network.

M.A.B., I.T. and J.H.K. acknowledge support from the State Research Agency (AEI) of the Ministry of Science and Innovation and the European Regional Development Fund (FEDER) under the grant with reference PID2019-105602GB-I00 and AYA2016-77237-C3-1-P, from IAC projects P/300624 and P/300724, financed by the Ministry of Science and Innovation, through the State Budget and by the Canary Islands Department of Economy, Knowledge and Employment, through the Regional Budget of the Autonomous Community, and from the Fundaci\'on BBVA under its 2017 programme of assistance to scientific research groups, for the project "Using machine-learning techniques to drag galaxies from the noise in deep imaging".   M.A.B. acknowledges support from the Severo  Ochoa  Excellence  scheme  (SEV-2015-0548).

\section*{Data Availability}
The data underlying this article were retrieved from the Hubble Space Telescope archive (HST Proposal 14643, PI: van Dokkum) and it provides the reduced/drizzled frames by the standard HST pipeline. The softwares and packages that are used in this work, \textit{SExtractor},  \textit{SWarp}, \textit{TinyTim} are publicly available. The catalogue of the Globular Cluster candidates around DF44 generated in this research is available in the article and in its online supplementary material. 

\bibliographystyle{mnras}
\bibliography{mnras_df44}

\begin{thebibliography}{}
\makeatletter
\relax
\def\mn@urlcharsother{\let\do\@makeother \do\$\do\&\do\#\do\^\do\_\do\%\do\~}
\def\mn@doi{\begingroup\mn@urlcharsother \@ifnextchar [ {\mn@doi@}
  {\mn@doi@[]}}
\def\mn@doi@[#1]#2{\def\@tempa{#1}\ifx\@tempa\@empty \href
  {http://dx.doi.org/#2} {doi:#2}\else \href {http://dx.doi.org/#2} {#1}\fi
  \endgroup}
\def\mn@eprint#1#2{\mn@eprint@#1:#2::\@nil}
\def\mn@eprint@arXiv#1{\href {http://arxiv.org/abs/#1} {{\tt arXiv:#1}}}
\def\mn@eprint@dblp#1{\href {http://dblp.uni-trier.de/rec/bibtex/#1.xml}
  {dblp:#1}}
\def\mn@eprint@#1:#2:#3:#4\@nil{\def\@tempa {#1}\def\@tempb {#2}\def\@tempc
  {#3}\ifx \@tempc \@empty \let \@tempc \@tempb \let \@tempb \@tempa \fi \ifx
  \@tempb \@empty \def\@tempb {arXiv}\fi \@ifundefined
  {mn@eprint@\@tempb}{\@tempb:\@tempc}{\expandafter \expandafter \csname
  mn@eprint@\@tempb\endcsname \expandafter{\@tempc}}}

\bibitem[\protect\citeauthoryear{{Amorisco} \& {Loeb}}{{Amorisco} \&
  {Loeb}}{2016}]{amorisco2016}
{Amorisco} N.~C.,  {Loeb} A.,  2016, \mn@doi [\mnras] {10.1093/mnrasl/slw055},
  \href {https://ui.adsabs.harvard.edu/abs/2016MNRAS.459L..51A} {459, L51}

\bibitem[\protect\citeauthoryear{{Amorisco}, {Monachesi}, {Agnello}  \&
  {White}}{{Amorisco} et~al.}{2018}]{amorisco2018}
{Amorisco} N.~C.,  {Monachesi} A.,  {Agnello} A.,   {White} S.~D.~M.,  2018,
  \mn@doi [\mnras] {10.1093/mnras/sty116}, \href
  {https://ui.adsabs.harvard.edu/abs/2018MNRAS.475.4235A} {475, 4235}

\bibitem[\protect\citeauthoryear{{Beasley} \& {Trujillo}}{{Beasley} \&
  {Trujillo}}{2016}]{beasley2016b}
{Beasley} M.~A.,  {Trujillo} I.,  2016, \mn@doi [\apj]
  {10.3847/0004-637X/830/1/23}, \href
  {https://ui.adsabs.harvard.edu/abs/2016ApJ...830...23B} {830, 23}

\bibitem[\protect\citeauthoryear{{Beasley}, {Romanowsky}, {Pota}, {Navarro},
  {Martinez Delgado}, {Neyer}  \& {Deich}}{{Beasley}
  et~al.}{2016}]{beasley2016}
{Beasley} M.~A.,  {Romanowsky} A.~J.,  {Pota} V.,  {Navarro} I.~M.,  {Martinez
  Delgado} D.,  {Neyer} F.,   {Deich} A.~L.,  2016, \mn@doi [\apjl]
  {10.3847/2041-8205/819/2/L20}, \href
  {https://ui.adsabs.harvard.edu/abs/2016ApJ...819L..20B} {819, L20}

\bibitem[\protect\citeauthoryear{{Bertin} \& {Arnouts}}{{Bertin} \&
  {Arnouts}}{1996}]{sex}
{Bertin} E.,  {Arnouts} S.,  1996, \mn@doi [\aaps] {10.1051/aas:1996164}, \href
  {https://ui.adsabs.harvard.edu/abs/1996A&AS..117..393B} {117, 393}

\bibitem[\protect\citeauthoryear{{Bertin}, {Mellier}, {Radovich}, {Missonnier},
  {Didelon}  \& {Morin}}{{Bertin} et~al.}{2002}]{swarp}
{Bertin} E.,  {Mellier} Y.,  {Radovich} M.,  {Missonnier} G.,  {Didelon} P.,
  {Morin} B.,  2002, {The TERAPIX Pipeline}.
p.~228

\bibitem[\protect\citeauthoryear{{Binggeli}}{{Binggeli}}{1994}]{Binggeli1994}
{Binggeli} B.,  1994, in European Southern Observatory Conference and Workshop
  Proceedings. p.~13

\bibitem[\protect\citeauthoryear{{Binggeli}, {Sandage}  \&
  {Tammann}}{{Binggeli} et~al.}{1985}]{Binggeli}
{Binggeli} B.,  {Sandage} A.,   {Tammann} G.~A.,  1985, \mn@doi [\aj]
  {10.1086/113874}, \href
  {https://ui.adsabs.harvard.edu/abs/1985AJ.....90.1681B} {90, 1681}

\bibitem[\protect\citeauthoryear{{Blakeslee} et~al.,}{{Blakeslee}
  et~al.}{2010}]{blak2010}
{Blakeslee} J.~P.,  et~al., 2010, \mn@doi [\apj] {10.1088/0004-637X/724/1/657},
  \href {https://ui.adsabs.harvard.edu/abs/2010ApJ...724..657B} {724, 657}

\bibitem[\protect\citeauthoryear{{Bogdan}}{{Bogdan}}{2020}]{akos}
{Bogdan} A.,  2020, arXiv e-prints, \href
  {https://ui.adsabs.harvard.edu/abs/2020arXiv200907846B} {p. arXiv:2009.07846}

\bibitem[\protect\citeauthoryear{{Chamba}, {Trujillo}  \& {Knapen}}{{Chamba}
  et~al.}{2020}]{chamba}
{Chamba} N.,  {Trujillo} I.,   {Knapen} J.~H.,  2020, \mn@doi [\aap]
  {10.1051/0004-6361/201936821}, \href
  {https://ui.adsabs.harvard.edu/abs/2020A&A...633L...3C} {633, L3}

\bibitem[\protect\citeauthoryear{{Collins} et~al.,}{{Collins}
  et~al.}{2013}]{collins2013}
{Collins} M. L.~M.,  et~al., 2013, \mn@doi [\apj]
  {10.1088/0004-637X/768/2/172}, \href
  {https://ui.adsabs.harvard.edu/abs/2013ApJ...768..172C} {768, 172}

\bibitem[\protect\citeauthoryear{{Dalcanton}, {Spergel}, {Gunn}, {Schmidt}  \&
  {Schneider}}{{Dalcanton} et~al.}{1997}]{Dalcanton}
{Dalcanton} J.~J.,  {Spergel} D.~N.,  {Gunn} J.~E.,  {Schmidt} M.,
  {Schneider} D.~P.,  1997, \mn@doi [\aj] {10.1086/118499}, \href
  {https://ui.adsabs.harvard.edu/abs/1997AJ....114..635D} {114, 635}

\bibitem[\protect\citeauthoryear{{Di Cintio}, {Brook}, {Dutton}, {Macci{\`o}},
  {Obreja}  \& {Dekel}}{{Di Cintio} et~al.}{2017}]{cintio2017}
{Di Cintio} A.,  {Brook} C.~B.,  {Dutton} A.~A.,  {Macci{\`o}} A.~V.,  {Obreja}
  A.,   {Dekel} A.,  2017, \mn@doi [\mnras] {10.1093/mnrasl/slw210}, \href
  {https://ui.adsabs.harvard.edu/abs/2017MNRAS.466L...1D} {466, L1}

\bibitem[\protect\citeauthoryear{{Emsellem} et~al.,}{{Emsellem}
  et~al.}{2019}]{Emsellem2019}
{Emsellem} E.,  et~al., 2019, \mn@doi [\aap] {10.1051/0004-6361/201834909},
  \href {https://ui.adsabs.harvard.edu/abs/2019A&A...625A..76E} {625, A76}

\bibitem[\protect\citeauthoryear{{Erwin}}{{Erwin}}{2015}]{2015ApJ...799..226E}
{Erwin} P.,  2015, \mn@doi [\apj] {10.1088/0004-637X/799/2/226}, \href
  {https://ui.adsabs.harvard.edu/abs/2015ApJ...799..226E} {799, 226}

\bibitem[\protect\citeauthoryear{{Fensch} et~al.,}{{Fensch}
  et~al.}{2019}]{fensch2019}
{Fensch} J.,  et~al., 2019, \mn@doi [\aap] {10.1051/0004-6361/201834911}, \href
  {https://ui.adsabs.harvard.edu/abs/2019A&A...625A..77F} {625, A77}

\bibitem[\protect\citeauthoryear{{Forbes}}{{Forbes}}{2017}]{forbes2017}
{Forbes} D.~A.,  2017, \mn@doi [\mnras] {10.1093/mnrasl/slx148}, \href
  {https://ui.adsabs.harvard.edu/abs/2017MNRAS.472L.104F} {472, L104}

\bibitem[\protect\citeauthoryear{{Forbes}, {Gannon}, {Couch}, {Iodice},
  {Spavone}, {Cantiello}, {Napolitano}  \& {Schipani}}{{Forbes}
  et~al.}{2019}]{udgngc4}
{Forbes} D.~A.,  {Gannon} J.,  {Couch} W.~J.,  {Iodice} E.,  {Spavone} M.,
  {Cantiello} M.,  {Napolitano} N.,   {Schipani} P.,  2019, \mn@doi [\aap]
  {10.1051/0004-6361/201935499}, \href
  {https://ui.adsabs.harvard.edu/abs/2019A&A...626A..66F} {626, A66}

\bibitem[\protect\citeauthoryear{{Gennaro, M., et al.}}{{Gennaro, M., et
  al.}}{2018}]{wfc3}
{Gennaro, M., et al.} 2018, {WFC3 Data Handbook, Version 4.0, (Baltimore:
  STScI)}

\bibitem[\protect\citeauthoryear{{Hanes}}{{Hanes}}{1977}]{gclf1}
{Hanes} D.~A.,  1977, \mn@doi [\mnras] {10.1093/mnras/180.3.309}, \href
  {https://ui.adsabs.harvard.edu/abs/1977MNRAS.180..309H} {180, 309}

\bibitem[\protect\citeauthoryear{{Harris}}{{Harris}}{2018}]{harris2018}
{Harris} W.~E.,  2018, \mn@doi [\aj] {10.3847/1538-3881/aaedb8}, \href
  {https://ui.adsabs.harvard.edu/abs/2018AJ....156..296H} {156, 296}

\bibitem[\protect\citeauthoryear{{Harris}, {Harris}  \& {Alessi}}{{Harris}
  et~al.}{2013}]{harris2013}
{Harris} W.~E.,  {Harris} G. L.~H.,   {Alessi} M.,  2013, \mn@doi [\apj]
  {10.1088/0004-637X/772/2/82}, \href
  {https://ui.adsabs.harvard.edu/abs/2013ApJ...772...82H} {772, 82}

\bibitem[\protect\citeauthoryear{{Harris}, {Blakeslee}  \& {Harris}}{{Harris}
  et~al.}{2017}]{harris2017}
{Harris} W.~E.,  {Blakeslee} J.~P.,   {Harris} G. L.~H.,  2017, \mn@doi [\apj]
  {10.3847/1538-4357/836/1/67}, \href
  {https://ui.adsabs.harvard.edu/abs/2017ApJ...836...67H} {836, 67}

\bibitem[\protect\citeauthoryear{{Hudson}, {Harris}  \& {Harris}}{{Hudson}
  et~al.}{2014}]{Hudson2014}
{Hudson} M.~J.,  {Harris} G.~L.,   {Harris} W.~E.,  2014, \mn@doi [\apjl]
  {10.1088/2041-8205/787/1/L5}, \href
  {https://ui.adsabs.harvard.edu/abs/2014ApJ...787L...5H} {787, L5}

\bibitem[\protect\citeauthoryear{{Impey} \& {Bothun}}{{Impey} \&
  {Bothun}}{1997}]{impey}
{Impey} C.,  {Bothun} G.,  1997, \mn@doi [\araa]
  {10.1146/annurev.astro.35.1.267}, \href
  {https://ui.adsabs.harvard.edu/abs/1997ARA&A..35..267I} {35, 267}

\bibitem[\protect\citeauthoryear{{Jord{\'a}n} et~al.,}{{Jord{\'a}n}
  et~al.}{2007}]{jordan2007}
{Jord{\'a}n} A.,  et~al., 2007, \mn@doi [\apjs] {10.1086/516840}, \href
  {https://ui.adsabs.harvard.edu/abs/2007ApJS..171..101J} {171, 101}

\bibitem[\protect\citeauthoryear{{Lim}, {Peng}, {C{\^o}t{\'e}}, {Sales}, {den
  Brok}, {Blakeslee}  \& {Guhathakurta}}{{Lim} et~al.}{2018}]{lim2018}
{Lim} S.,  {Peng} E.~W.,  {C{\^o}t{\'e}} P.,  {Sales} L.~V.,  {den Brok} M.,
  {Blakeslee} J.~P.,   {Guhathakurta} P.,  2018, \mn@doi [\apj]
  {10.3847/1538-4357/aacb81}, \href
  {https://ui.adsabs.harvard.edu/abs/2018ApJ...862...82L} {862, 82}

\bibitem[\protect\citeauthoryear{{Mancera Pi{\~n}a}, {Aguerri}, {Peletier},
  {Venhola}, {Trager}  \& {Choque Challapa}}{{Mancera Pi{\~n}a}
  et~al.}{2019}]{pavel2019}
{Mancera Pi{\~n}a} P.~E.,  {Aguerri} J.~A.~L.,  {Peletier} R.~F.,  {Venhola}
  A.,  {Trager} S.,   {Choque Challapa} N.,  2019, \mn@doi [\mnras]
  {10.1093/mnras/stz238}, \href
  {https://ui.adsabs.harvard.edu/abs/2019MNRAS.485.1036M} {485, 1036}

\bibitem[\protect\citeauthoryear{{Mancera Pi{\~n}a} et~al.,}{{Mancera Pi{\~n}a}
  et~al.}{2020}]{pavel2020}
{Mancera Pi{\~n}a} P.~E.,  et~al., 2020, arXiv e-prints, \href
  {https://ui.adsabs.harvard.edu/abs/2020arXiv200414392M} {p. arXiv:2004.14392}

\bibitem[\protect\citeauthoryear{{M{\"u}ller} et~al.,}{{M{\"u}ller}
  et~al.}{2020}]{oliver2020}
{M{\"u}ller} O.,  et~al., 2020, arXiv e-prints, \href
  {https://ui.adsabs.harvard.edu/abs/2020arXiv200604606M} {p. arXiv:2006.04606}

\bibitem[\protect\citeauthoryear{{Peng} \& {Lim}}{{Peng} \&
  {Lim}}{2016}]{peng2016}
{Peng} E.~W.,  {Lim} S.,  2016, \mn@doi [\apjl] {10.3847/2041-8205/822/2/L31},
  \href {https://ui.adsabs.harvard.edu/abs/2016ApJ...822L..31P} {822, L31}

\bibitem[\protect\citeauthoryear{{Peng} et~al.,}{{Peng}
  et~al.}{2011}]{peng2011}
{Peng} E.~W.,  et~al., 2011, \mn@doi [\apj] {10.1088/0004-637X/730/1/23}, \href
  {https://ui.adsabs.harvard.edu/abs/2011ApJ...730...23P} {730, 23}

\bibitem[\protect\citeauthoryear{{Prole} et~al.,}{{Prole}
  et~al.}{2019}]{udgngc3}
{Prole} D.~J.,  et~al., 2019, \mn@doi [\mnras] {10.1093/mnras/stz326}, \href
  {https://ui.adsabs.harvard.edu/abs/2019MNRAS.484.4865P} {484, 4865}

\bibitem[\protect\citeauthoryear{{Rejkuba}}{{Rejkuba}}{2012}]{rejkuba2012}
{Rejkuba} M.,  2012, \mn@doi [\apss] {10.1007/s10509-012-0986-9}, \href
  {https://ui.adsabs.harvard.edu/abs/2012Ap&SS.341..195R} {341, 195}

\bibitem[\protect\citeauthoryear{{Rom{\'a}n} \& {Trujillo}}{{Rom{\'a}n} \&
  {Trujillo}}{2017}]{javier}
{Rom{\'a}n} J.,  {Trujillo} I.,  2017, \mn@doi [\mnras] {10.1093/mnras/stx694},
  \href {https://ui.adsabs.harvard.edu/abs/2017MNRAS.468.4039R} {468, 4039}

\bibitem[\protect\citeauthoryear{{Rong} et~al.,}{{Rong}
  et~al.}{2019}]{Rong2019}
{Rong} Y.,  et~al., 2019, arXiv e-prints, \href
  {https://ui.adsabs.harvard.edu/abs/2019arXiv190710079R} {p. arXiv:1907.10079}

\bibitem[\protect\citeauthoryear{{Ruiz-Lara} et~al.,}{{Ruiz-Lara}
  et~al.}{2018}]{ruizlara2018}
{Ruiz-Lara} T.,  et~al., 2018, \mn@doi [\mnras] {10.1093/mnras/sty1112}, \href
  {https://ui.adsabs.harvard.edu/abs/2018MNRAS.478.2034R} {478, 2034}

\bibitem[\protect\citeauthoryear{{Sandage} \& {Binggeli}}{{Sandage} \&
  {Binggeli}}{1984}]{Sandage1984}
{Sandage} A.,  {Binggeli} B.,  1984, \mn@doi [\aj] {10.1086/113588}, \href
  {https://ui.adsabs.harvard.edu/abs/1984AJ.....89..919S} {89, 919}

\bibitem[\protect\citeauthoryear{{Secker} \& {Harris}}{{Secker} \&
  {Harris}}{1993}]{1993AJ....105.1358S}
{Secker} J.,  {Harris} W.~E.,  1993, \mn@doi [\aj] {10.1086/116515}, \href
  {https://ui.adsabs.harvard.edu/abs/1993AJ....105.1358S} {105, 1358}

\bibitem[\protect\citeauthoryear{{Spitler} \& {Forbes}}{{Spitler} \&
  {Forbes}}{2009}]{Spitler2009}
{Spitler} L.~R.,  {Forbes} D.~A.,  2009, \mn@doi [\mnras]
  {10.1111/j.1745-3933.2008.00567.x}, \href
  {https://ui.adsabs.harvard.edu/abs/2009MNRAS.392L...1S} {392, L1}

\bibitem[\protect\citeauthoryear{{Trujillo}, {Graham}  \& {Caon}}{{Trujillo}
  et~al.}{2001}]{2001MNRAS.326..869T}
{Trujillo} I.,  {Graham} A.~W.,   {Caon} N.,  2001, \mn@doi [\mnras]
  {10.1046/j.1365-8711.2001.04471.x}, \href
  {https://ui.adsabs.harvard.edu/abs/2001MNRAS.326..869T} {326, 869}

\bibitem[\protect\citeauthoryear{{Trujillo} et~al.,}{{Trujillo}
  et~al.}{2019}]{2019MNRAS.486.1192T}
{Trujillo} I.,  et~al., 2019, \mn@doi [\mnras] {10.1093/mnras/stz771}, \href
  {https://ui.adsabs.harvard.edu/abs/2019MNRAS.486.1192T} {486, 1192}

\bibitem[\protect\citeauthoryear{{Trujillo}, {Chamba}  \& {Knapen}}{{Trujillo}
  et~al.}{2020}]{nacho2020}
{Trujillo} I.,  {Chamba} N.,   {Knapen} J.~H.,  2020, \mn@doi [\mnras]
  {10.1093/mnras/staa236}, \href
  {https://ui.adsabs.harvard.edu/abs/2020MNRAS.tmp..226T} {p.~226}

\bibitem[\protect\citeauthoryear{{Vazdekis}, {Koleva}, {Ricciardelli},
  {R{\"o}ck}  \& {Falc{\'o}n-Barroso}}{{Vazdekis} et~al.}{2016}]{vazdekis2016}
{Vazdekis} A.,  {Koleva} M.,  {Ricciardelli} E.,  {R{\"o}ck} B.,
  {Falc{\'o}n-Barroso} J.,  2016, \mn@doi [\mnras] {10.1093/mnras/stw2231},
  \href {https://ui.adsabs.harvard.edu/abs/2016MNRAS.463.3409V} {463, 3409}

\bibitem[\protect\citeauthoryear{{Venhola} et~al.,}{{Venhola}
  et~al.}{2017}]{venhola2017}
{Venhola} A.,  et~al., 2017, \mn@doi [\aap] {10.1051/0004-6361/201730696},
  \href {https://ui.adsabs.harvard.edu/abs/2017A&A...608A.142V} {608, A142}

\bibitem[\protect\citeauthoryear{{van Dokkum}, {Abraham}, {Merritt}, {Zhang},
  {Geha}  \& {Conroy}}{{van Dokkum} et~al.}{2015a}]{vd15}
{van Dokkum} P.~G.,  {Abraham} R.,  {Merritt} A.,  {Zhang} J.,  {Geha} M.,
  {Conroy} C.,  2015a, \mn@doi [\apjl] {10.1088/2041-8205/798/2/L45}, \href
  {https://ui.adsabs.harvard.edu/abs/2015ApJ...798L..45V} {798, L45}

\bibitem[\protect\citeauthoryear{{van Dokkum} et~al.,}{{van Dokkum}
  et~al.}{2015b}]{vd15b}
{van Dokkum} P.~G.,  et~al., 2015b, \mn@doi [\apjl]
  {10.1088/2041-8205/804/1/L26}, \href
  {https://ui.adsabs.harvard.edu/abs/2015ApJ...804L..26V} {804, L26}

\bibitem[\protect\citeauthoryear{{van Dokkum} et~al.,}{{van Dokkum}
  et~al.}{2016}]{vd16}
{van Dokkum} P.,  et~al., 2016, \mn@doi [\apjl] {10.3847/2041-8205/828/1/L6},
  \href {https://ui.adsabs.harvard.edu/abs/2016ApJ...828L...6V} {828, L6}

\bibitem[\protect\citeauthoryear{{van Dokkum} et~al.,}{{van Dokkum}
  et~al.}{2017}]{vd17}
{van Dokkum} P.,  et~al., 2017, \mn@doi [\apjl] {10.3847/2041-8213/aa7ca2},
  \href {https://ui.adsabs.harvard.edu/abs/2017ApJ...844L..11V} {844, L11}

\bibitem[\protect\citeauthoryear{{van Dokkum} et~al.,}{{van Dokkum}
  et~al.}{2018}]{2018Natur.555..629V}
{van Dokkum} P.,  et~al., 2018, \mn@doi [\nat] {10.1038/nature25767}, \href
  {https://ui.adsabs.harvard.edu/abs/2018Natur.555..629V} {555, 629}

\bibitem[\protect\citeauthoryear{{van Dokkum} et~al.,}{{van Dokkum}
  et~al.}{2019}]{vd19}
{van Dokkum} P.,  et~al., 2019, \mn@doi [\apj] {10.3847/1538-4357/ab2914},
  \href {https://ui.adsabs.harvard.edu/abs/2019ApJ...880...91V} {880, 91}

\makeatother
\end{thebibliography}

\appendix

\section{Likelihood of the S\'ersic function}

In this Appendix we describe the Likelihood function associated to a distribution of globular clusters following a spatial configuration dictated by a S\'ersic law. We assume the globular clusters are located around the host with an elliptical symmetry. We consider the center of the distribution (x$_0$, y$_0$), the position angle ($\alpha$) and the axis ratio q to be known. With such assumption, the distance of each globular cluster to the center of the distribution can be described by:

\begin{equation}
R_i=\left(x^2_{GC,i}+\frac{y^2_{GC,i}}{q^2}\right)^{1/2}    
\end{equation}

with 

\begin{eqnarray}
x_{GC,i}=(x_i-x_0)\cos{\alpha}+(y_i-y_0)\sin{\alpha} \\
y_{GC,i}=-(x_i-x_0)\sin{\alpha}+(y_i-y_0)\cos{\alpha}
\end{eqnarray}

The S\'ersic distribution probability density function is:

\begin{equation}
    S(R)=\frac{b_n^{2n} e^{-b_n\left(R/R_{GC}\right)^{1/n}}}{2 \pi q R_{GC}^2  n \Gamma(2n)}
\end{equation}
with n the S\'ersic index and R$_{GC}$ the effective radius of the globular cluster distribution.  b$_n$ satisfies $\gamma$(2n,b$_n$)=1/2$\Gamma$(2n) with $\Gamma$ and $\gamma$ the complete and incomplete gamma functions respectively.

In our case, the parameters of the Likelihood function are $\theta$=(R$_{GC}$,n). For a dataset of size N (i.e. for a number of globular clusters equal to N), the probability of our observation can be represented by L(R$_1$,R$_2$,...,R$_N$|$\theta$). From probability theory, we know that the probability of multiple independent events all happening is termed the joint probability, therefore we can write our Likelihood as:

\begin{equation}
   L(R_1,R_2,...,R_N|\theta)=\prod_{i=1}^{N} L(R_i|\theta)
\end{equation}

For convenience, we work using the log Likelihood function:

\begin{equation}
LL(R_1,R_2,...,R_N|\theta)=\sum_{i=1}^{N} \ln{\frac{b_n^{2n} e^{-b_n\left(R_i/R_{GC}\right)^{1/n}}}{2 \pi q R_{GC}^2  n \Gamma(2n)}}
\end{equation}

A bit of algebra shows that:

\begin{eqnarray}
\label{EqLL}
    LL(R_1,R_2,...,R_N|\theta) & = & -2N\ln{R_{GC}}+N\ln{\left(\frac{b_n^{2n}}{n\Gamma{(2n)}}\right)}-N\ln{(2\pi)}-
    \nonumber \\
  & & {} -N\ln{q}-\frac{b_n}{R_{GC}^{1/n}}\sum_{i=1}^{N}R_i^{1/n}
\end{eqnarray}

\subsection{Maximum Likelihood Estimation for R$_{GC}$}

In the case of R$_{GC}$, the maximum likelihood value can be estimated analytically relatively simply.  To get this value we calculate $\partial$LL/$\partial$R$_{GC}$=0. Using Eq. \ref{EqLL}, we get:

\begin{equation}
 \frac{\partial LL}{\partial R_{GC}} = -2N\frac{\partial}{\partial R_{GC}}\ln{R_{GC}}-\frac{\partial}{\partial R_{GC}}\left( \frac{b_n}{R_{GC}^{1/n}}\sum_{i=1}^{N}R_i^{1/n}\right)
\end{equation}

From that equation we obtain:

\begin{equation}
\frac{\partial LL}{\partial R_{GC}} = -2N\frac{1}{R_{GC}} + \frac{b_n}{n}\frac{1}{R_{GC}}\frac{1}{R_{GC}^{1/n}}\sum_{i=1}^{N}R_i^{1/n}
\end{equation}

Finally, applying $\partial$LL/$\partial$R$_{GC}$=0, we get the following solution of the maximum likelihood value for R$_{GC}$:

\begin{equation}
R_{GC}=\frac{b_n^n}{(2n)^n}\frac{1}{N}\sum_{i=1}^{N}R_i^{1/n}
\label{rgcequation}
\end{equation}

\bsp
\label{lastpage}

\end{document}